\pdfoutput=1
\documentclass[11pt,twoside,a4paper,cmspaper,final,collab]{cms-tdr}
\def\svnVersion{2607d03}\def\svnDate{2022/09/12}\def\cmsCernNoTag{CERN-EP-2020-181}\def\cmsCernDate{\today}\def\cmsMessage{Published in Physics Letters B as \href{http://dx.doi.org/10.1016/j.physletb.2022.137397}{\doi{10.1016/j.physletb.2022.137397}.}}
\begin{document}\cmsNoteHeader{HIN-18-005}

\newlength\cmsFigWidth
\ifthenelse{\boolean{cms@external}}{\setlength\cmsFigWidth{0.85\columnwidth}}{\setlength\cmsFigWidth{0.4\textwidth}}
\ifthenelse{\boolean{cms@external}}{\providecommand{\cmsLeft}{top\xspace}}{\providecommand{\cmsLeft}{left\xspace}}
\ifthenelse{\boolean{cms@external}}{\providecommand{\cmsRight}{bottom\xspace}}{\providecommand{\cmsRight}{right\xspace}}
\newlength\cmsTabSkip\setlength{\cmsTabSkip}{1ex}

\newcommand{\ptmu}{\ensuremath{p^{\PGm}_{\mathrm{T}}}\xspace}
\newcommand{\ptmumu}{\ensuremath{p^{\PGm\PGm}_{\mathrm{T}}}\xspace}
\newcommand{\ptUps}{\ensuremath{p^{\PGU}_{\mathrm{T}}}\xspace}
\newcommand{\rap}{\ensuremath{y_{\mathrm{CM}}}\xspace}
\newcommand{\rapmumu}{\ensuremath{y^{\PGm\PGm}_{\mathrm{CM}}}\xspace}
\newcommand{\rapUps}{\ensuremath{y^{\PGU}_{\mathrm{CM}}}\xspace}
\newcommand{\ylab}{\ensuremath{y_{\text{lab}}}\xspace}
\newcommand{\ylabmumu}{\ensuremath{y^{\PGm\PGm}_{\text{lab}}}\xspace}
\newcommand{\ylabUps}{\ensuremath{y^{\PGU}_{\text{lab}}}\xspace}
\newcommand{\etalab}{\ensuremath{\eta_{\text{lab}}}\xspace}
\newcommand{\etalabmu}{\ensuremath{\eta^{\PGm}_{\text{lab}}}\xspace}

\newcommand{\sqrts}{\ensuremath{\sqrt{s}}\xspace}

\newcommand{\YnS}{\PGUP{\text{\ensuremath{n}S}}}
\newcommand{\pp}{\ensuremath{\Pp\Pp}\xspace}
\newcommand{\PbPb}{\ensuremath{\mathrm{PbPb}}\xspace}
\newcommand{\pPb}{\ensuremath{\Pp\mathrm{Pb}}\xspace}
\newcommand{\Pbp}{\ensuremath{\mathrm{Pb}\Pp}\xspace}
\newcommand{\muplusmuminus}{\ensuremath{\PGmp\PGmm}\xspace}
\newcommand{\RAA}{\ensuremath{R_{\mathrm{AA}}}\xspace}
\newcommand{\RpPb}{\ensuremath{R_{\Pp\mathrm{Pb}}}\xspace}
\newcommand{\RFB}{\ensuremath{R_{\mathrm{FB}}}\xspace}

\newcommand{\Ntracks}{\ensuremath{N_{\text{tracks}}}\xspace}

\ifthenelse{\boolean{cms@external}}{\providecommand{\cmsLeft}{upper\xspace}}{\providecommand{\cmsLeft}{left\xspace}}
\ifthenelse{\boolean{cms@external}}{\providecommand{\cmsRight}{lower\xspace}}{\providecommand{\cmsRight}{right\xspace}}

\cmsNoteHeader{HIN-18-xxx}

\title{Nuclear modification of \texorpdfstring{\PgU}{Upsilon} states in \texorpdfstring{\pPb}{pPb} collisions at \texorpdfstring{$\sqrtsNN = 5.02\TeV$}{sqrt(sNN) = 5.02 TeV}}

\date{\today}

\abstract{
Production cross sections of \PgUa, \PgUb, and \PgUc states decaying into \muplusmuminus in proton-lead (\pPb) collisions are reported using data collected by the CMS experiment at $\sqrtsNN = 5.02\TeV$. A comparison is made with corresponding cross sections obtained with \pp data measured at the same collision energy and scaled by the Pb nucleus mass number. The nuclear modification factor for \PgUa is found to be $\RpPb(\PgUa) = 0.806 \pm 0.024 \stat \pm 0.059 \syst$. Similar results for the excited states indicate a sequential suppression pattern, such that $\RpPb(\PgUa) > \RpPb(\PgUb) > \RpPb(\PgUc)$. The suppression of all states is much less pronounced in \pPb than in \PbPb collisions, and independent of transverse momentum \ptUps and center-of-mass rapidity \rapUps of the individual \PgU state in the studied range $\ptUps < 30\GeVc$ and $\abs{\rapUps} < 1.93$. Models that incorporate final-state effects of bottomonia in pPb collisions are in better agreement with the data than those which only assume initial-state modifications. 
}

\hypersetup{
pdfauthor={CMS Collaboration},
pdftitle={Nuclear modification of Upsilon states in pPb collisions at sqrt(s[NN]) = 5.02 TeV},
pdfsubject={CMS, bottomonium, Upsilons},
pdfkeywords={CMS,  bottomonium, quarkonium suppression, quark-gluon plasma, heavy ion collisions}}

\maketitle

\section{Introduction}
Properties of the color-deconfined quark-gluon plasma (QGP) created in high-energy collisions of heavy nuclei can be studied using heavy-quark resonances produced by initial hard scatterings~\cite{matsui,harris,gunion,kim,emerick,Rothkopf:2016vsn}.
Yields of various quarkonium states, which have a short formation time in their rest frames and can typically escape the QGP before they decay, 
encode information on the evolution of the plasma starting from its early stages~\cite{matsui,harris,andronic,digal,kim,Rothkopf:2016vsn,BurnierKaczmarekRothkopf}.
Debye screening and gluo-dissociation~\cite{laine,blaizot,chen,chu,Brambilla:2008cx} in the QGP produced in lead-lead (\PbPb) collisions are understood to modify yields of quarkonium states hierarchically,
according to their binding energies. Each state dissociates when a high enough temperature is reached in the QGP~\cite{kim,Rothkopf:2016vsn,burnier,emerick,BurnierKaczmarekRothkopf}.
To interpret the quarkonium-state suppression patterns observed in heavy ion collisions as signals of color deconfinement in the hot plasma, it is essential to understand initial state and final state "cold nuclear matter" (CNM) effects.
In this context, initial state refers to the partons in the relevant quantum chromodynamics process that stem from the colliding proton (\Pp) or nucleus and scatter to produce a heavy quark pair but before it hadronizes into a quarkonium state.
Examples of CNM effects that have been discussed in pA collisions include shadowing of the parton distribution functions in the nucleus (initial state)~\cite{VogtComp}, energy loss in the nucleus (initial and final states)~\cite{ArleoComp}, and interactions with hadronic comovers (final state)~\cite{FerreiroComp}. For a recent review, see Ref.~\cite{andronic}.
Traditionally, all modifications observed in \pPb collisions were assumed to be due to CNM effects. However, it is worth noting that this assumption has been questioned given recent evidence of collective behavior in \pp and \pPb collisions with the highest amount of emitted particles, further referred to as high-activity~\cite{CMS:2010ifv,Chatrchyan:2013cor,ALICE:2016fzo,Khachatryan:2015cor,ATLAS:2012cix,ATLAS:2016yzd,ATLAS:2019xqc}.
This might be explained by assuming the formation of a QGP-like medium~\cite{LHC_smallsystem}.

Bottomonia serve as particularly powerful probes for studying the QGP, since their high masses require that their production be dominated by initial hard scattering of partons in the collisions~\cite{CNMvogt,andronic,strickland,ArleoDetailed,RothkopfKrouppaStrickland}. When compared to charmonia, their yields are considerably less modified by regeneration or recombination in the QGP~\cite{ChenZhao,rapp,RothkopfKrouppaStrickland}.
Measurements by the CMS experiment showing sequential modification of \YnS (where $n=1,2,3$) decaying via the dimuon channel in \PbPb compared with \pp collisions at a nucleon-nucleon center-of-mass energy of $\sqrtsNN = 2.76\TeV$~\cite{CMS:2011all,HIN-15-001} and $5.02\TeV$~\cite{HIN-16-008,HIN-16-023} were used to infer model-dependent~\cite{rapp,KrouppaStrickland} QGP temperatures. 
This effect is consistent with models that incorporate sequential suppression due to color screening~\cite{emerick,digal,RothkopfKrouppaStrickland}.
Similar measurements of \YnS production in \pPb collisions can help to disentangle hot and cold nuclear matter effects and to investigate various CNM mechanisms.

Nuclear modification factors \RpPb are ratios of particle production cross sections in \pPb collisions over the corresponding cross sections in \pp collisions scaled to account for the number of nucleons in the Pb nucleus.
The \RpPb values quantify the modification of hard probe production in \pPb collisions due to the nuclear environment created by a single lead nucleus in the initial state.
In this analysis, these factors are determined for \YnS under the assumption that the cross sections scale as $\sigma_{\pPb} = A \sigma_{\pp}$, where $A$ is the mass number of Pb.
With this assumption, also known as the A-scaling hypothesis, values of \RpPb different from unity indicate modifications that go beyond simple superposition of binary nucleon-nucleon collisions.
These \RpPb values, together with measurements of the nuclear modification factors \RAA in \PbPb collisions~\cite{HIN-16-023}, can be used to investigate the relative contributions of hot and cold nuclear matter effects.

Since \pPb collisions create an imbalance of nuclear matter in the proton-going (forward rapidity) and lead-going (backward rapidity) directions, they can be used to investigate 
differences in CNM effects in these regions of varying nuclear matter density within the same collision system. 
In the charmonium sector, CMS has found hints of differences in the level of suppression between the excited and ground state in the lead-going region~\cite{HIN-14-009,HIN-16-015}.
One CNM modification mechanism that relies on the abundance of nuclear matter is dissociation by interaction with comoving particles,
where the cross section of interaction increases with particle multiplicity in the rapidity region of the produced \PgU meson~\cite{FerreiroComp,FerreiroCharm15}.
This is quantified by measuring the forward-backward production ratios \RFB of \PgU states in \pPb collisions.

The LHCb~\cite{LHCbRpA} and ALICE~\cite{ALICERpA} Collaborations reported measurements of the \YnS/\PgUa yield ratios (LHCb for $n=2$ and 3; ALICE for $n=2$), along with \RpPb and \RFB for \PgUa in \pPb collisions at $\sqrtsNN = 5.02\TeV$ using \PgU mesons detected in the forward rapidity region.
In those studies, the proton reference was obtained by interpolating results from event samples collected at other collision energies, \ie, 2.76, 7, and 8\TeV. 
In the midrapidity region, the ATLAS Collaboration studied bottomonia in \pPb collisions
using same-energy \pp reference data~\cite{ATLASRpA}, reporting \YnS/\PgUa (for $n=2$ and 3), as well as \PgUa yields self-normalized to their activity-integrated values, and \RpPb(\PgUa).
The CMS Collaboration previously reported the \YnS/\PgUa (for $n=2$ and 3) yield ratios versus event activity in the \pPb system at $\sqrtsNN = 5.02\TeV$~\cite{HIN-13-003}, as well as in \pp collisions at $\sqrts=2.76\TeV$~\cite{HIN-13-003} and 7\TeV~\cite{CMS:2020fae}.
More recently, the LHCb~\cite{LHCbRpA2} and ALICE~\cite{ALICERpA2} Collaborations measured \RpPb and \RFB for both \PgUa and \PgUb at the higher energy $\sqrtsNN = 8.16\TeV$, using \pp reference data interpolated from measurements at $\sqrts = 2.76,$ 7, 8, and 13\TeV. 
Data for \PbPb collisions are not available at 8.16\TeV for direct comparison.
These bottomonium measurements in \pPb have focused on the ground state and indicate that the level of suppression is consistent with that expected from shadowing calculations, but they provide little information on the behavior of the excited states.

In this Letter, we analyze \pPb and \pp collision data from the CERN LHC collected with the CMS detector at the same nucleon-nucleon center-of-mass (CM) energy of $\sqrtsNN = 5.02\TeV$. The yields of \YnS mesons are measured using their decay to two muons.
By comparing the yields measured in the two colliding systems, the \RpPb and \RFB factors are determined including all bottomonium states for the first time.
Because models that incorporate final-state CNM effects are the only ones to predict different modifications for the excited states,
these measurements for the ordering of excited state \RpPb values may reveal these types of final-state mechanisms.
Ordered suppression could arise from various causes --- \eg, the size of the states, their cross section with potential comovers, or their binding energy.
These results are compared with measurements of the \YnS nuclear modification factors \RAA in \PbPb collisions~\cite{HIN-16-023} using \PbPb data also collected at 5.02\TeV with the CMS detector, allowing a model-dependent comparison of bottomonia in hot and cold nuclear matter.

\section{The CMS detector}
The central feature of the CMS apparatus is a superconducting solenoid of 6\unit{m} internal diameter, providing a magnetic field of 3.8\unit{T}.
Within the solenoid volume are a silicon pixel and strip tracker, a lead tungstate crystal electromagnetic calorimeter, and a brass and scintillator hadron calorimeter, each composed of a barrel and two endcap sections.
Muons are detected in the range $\abs{\etalab}<2.4$ in gas-ionization detectors embedded in the steel flux-return yoke outside the solenoid.
In the barrel region $\abs{\etalab}<1.2$ muon detection planes are based on drift tube technology, while the endcap region $0.9<\abs{\etalab}<2.4$ uses cathode strip chambers. Resistive plate chambers provide additional muon detection capability in the range $\abs{\etalab}<1.6$.
Matching muons to tracks measured in the silicon tracker leads to a relative transverse momentum \pt\ resolution on the order of 1\%
for a typical muon used in this analysis~\cite{Chatrchyan:2012xi}. 
In addition, two steel and quartz-fiber hadron forward calorimeters cover the range $2.9<\abs{\etalab}<5.2$.
A detailed description of the CMS detector, together with a definition of the coordinate system used and the relevant kinematic variables, can be found in Ref.~\cite{Chatrchyan:2008zzk}. 

A two-tiered system is used to select collision events of interest from the detector. The first level (L1), composed of custom hardware processors, uses information from the calorimeters and muon detectors to select events at a rate of around 100\unit{kHz} within a fixed latency of about 4\mus~\cite{Sirunyan:2020zal}.
The second level, known as the high-level trigger (HLT), consists of a farm of processors running a version of the full event reconstruction software optimized for fast processing, and reduces the event rate to around 1\unit{kHz} before data storage~\cite{Khachatryan:2016bia}.

\section{Data selection and simulated samples}
The events used for this analysis are selected using the trigger systems described above, requiring two muon candidates in the muon detectors with no explicit cuts in muon transverse momentum, \ptmu, or muon pseudorapidity measured in the laboratory, $\etalabmu$.
The event samples used in this analysis correspond to integrated luminosities of $28.0 \pm 0.6\pbinv$ and $34.6 \pm 1.2\nbinv$ for \pp ~\cite{pp_Lumi} and \pPb ~\cite{pPb_Lumi} collisions, respectively.
The uncertainties in the integrated luminosity determination are considered as a global uncertainty in all results.
All recorded \pPb events are required to have an energy deposit above 3\GeV in the hadron forward calorimeters on each side of the interaction point in order to suppress background from ultra-peripheral collisions and beam-gas events, while having a high efficiency for the selection of beam-beam hadronic collisions.

In the case of \pPb collisions, the value of the integrated luminosity represents the combined luminosity of collisions with proton and lead beams traveling in either direction.
While in the symmetric \pp and \PbPb collision systems the CM and laboratory (lab) reference frames coincide, in the case of \pPb collisions the difference between the energy-per-nucleon of the two beams induces a shift between the two frames.
For \pPb collisions at $\sqrtsNN = 5.02\TeV$, the rapidity y is 
shifted in the CM frame by $\delta y = 0.465$ compared to the lab frame. 
The rapidity range of the reconstructed dimuons in the lab frame $\abs{\ylabmumu}<2.4$ corresponds to a CM frame rapidity range of either $-2.87 < \rapmumu < 1.93$ (\Pbp) or $-1.93 < \rapmumu < 2.87$ (\pPb), depending on the direction of the proton beam.
In order to minimize the influence of asymmetric detector conditions, data are taken with both beam directions and then combined by inverting the rapidity of one of the datasets.

For both \pp and \pPb data, we select events with muon candidates in the kinematic range $\ptmu > 4\GeVc$, $\abs{\etalabmu} < 2.4$. The muon tracks are required to have at least 6 hits in the silicon tracker, at least one hit in the silicon pixel detector, and match with at least one segment in any detection plane of the muon system. 
The distance of the track from the closest primary vertex~\cite{primary_vertex} must be less than 20\cm in the longitudinal direction and 0.3\cm in the transverse direction. 
When forming a muon pair, each of the two muons is required to match the hardware trigger that prompted recording of the event
and to originate from a common vertex with a $\chi^2$ probability larger than 1\%, as obtained by a Kalman vertex filter algorithm~\cite{kalman}. 
For \pPb data, an additional filter is used to remove events that contain multiple interactions per bunch crossing (pileup)~\cite{HIN-14-009}.
This filter reduces the fraction of pileup events from 3\% to less than 0.2\%, and reduces the effective luminosity of \pPb collisions by 4.1\% compared to the numbers noted above.

Dedicated Monte Carlo (MC) simulations of collision data are used to validate fitting techniques and to correct the extracted \YnS yields for losses due to finite detector acceptance and efficiency. 
Simulated samples are independently generated for the \PgUa, \PgUb, and \PgUc mesons, in \pp collisions using \PYTHIA 8.209~\cite{Sjostrand:2014zea}, assuming no polarization based on measurements at the LHC~\cite{CMS:2012bpf,LHCb:2017scf}.
To simulate \pPb collisions, the rapidities of all particles in the generated \pp events are boosted by $\delta y = 0.465$ in the Pb-going direction to mimic the \rap shift in data.
The CMS detector response is simulated using \GEANTfour~\cite{Agostinelli:2002hh}. 
The reconstructed \ptUps\ distributions of the simulated \PgU states are weighted using a fit to the ratio of the \ptUps\ spectra in data and simulation.
The rapidity distributions in simulation are consistent with those in data.

\section{Analysis}
\subsection{Signal extraction}
\begin{figure}[htb!]
\begin{centering}
\includegraphics[width=0.45\textwidth]{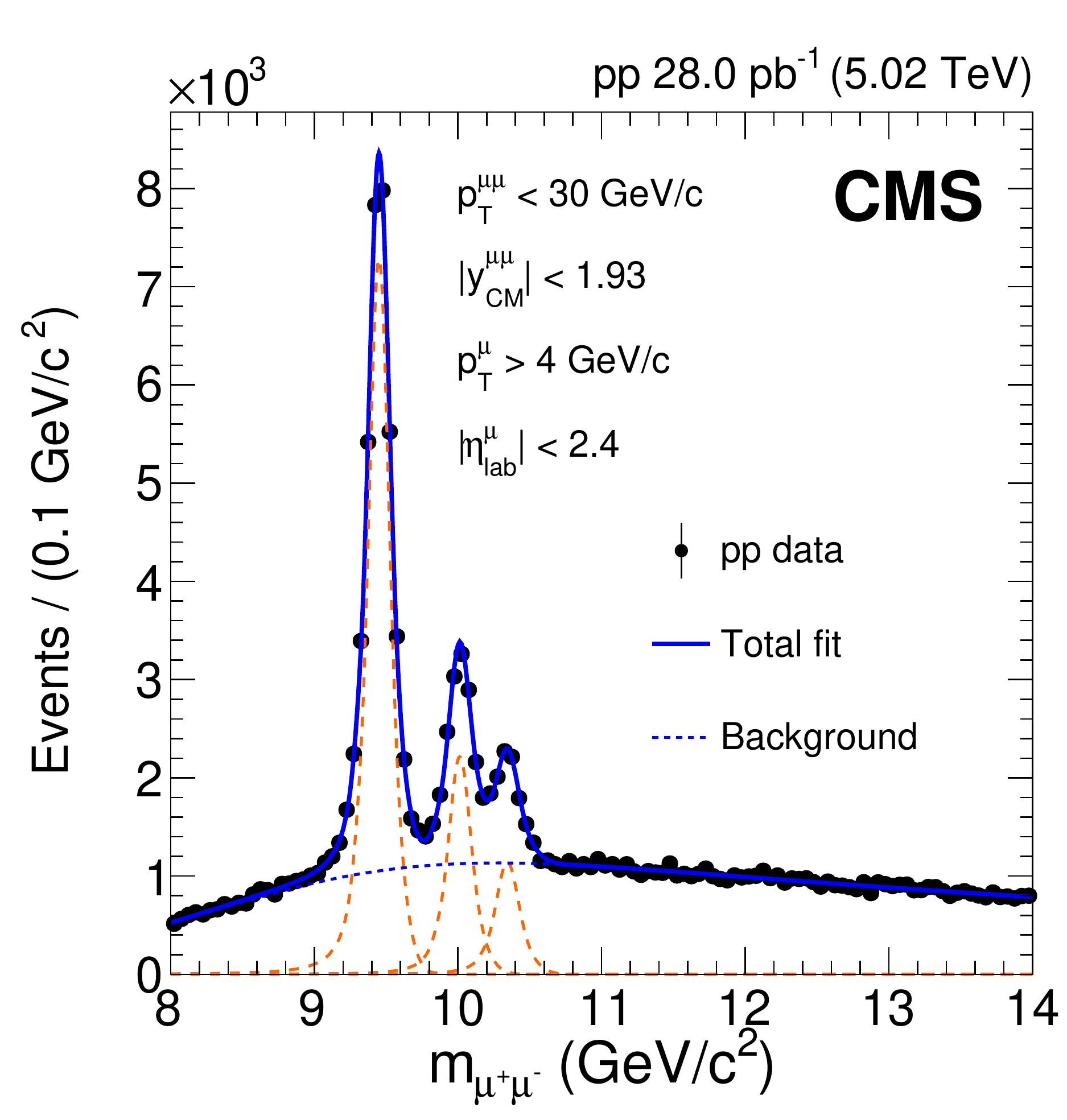}
\includegraphics[width=0.45\textwidth]{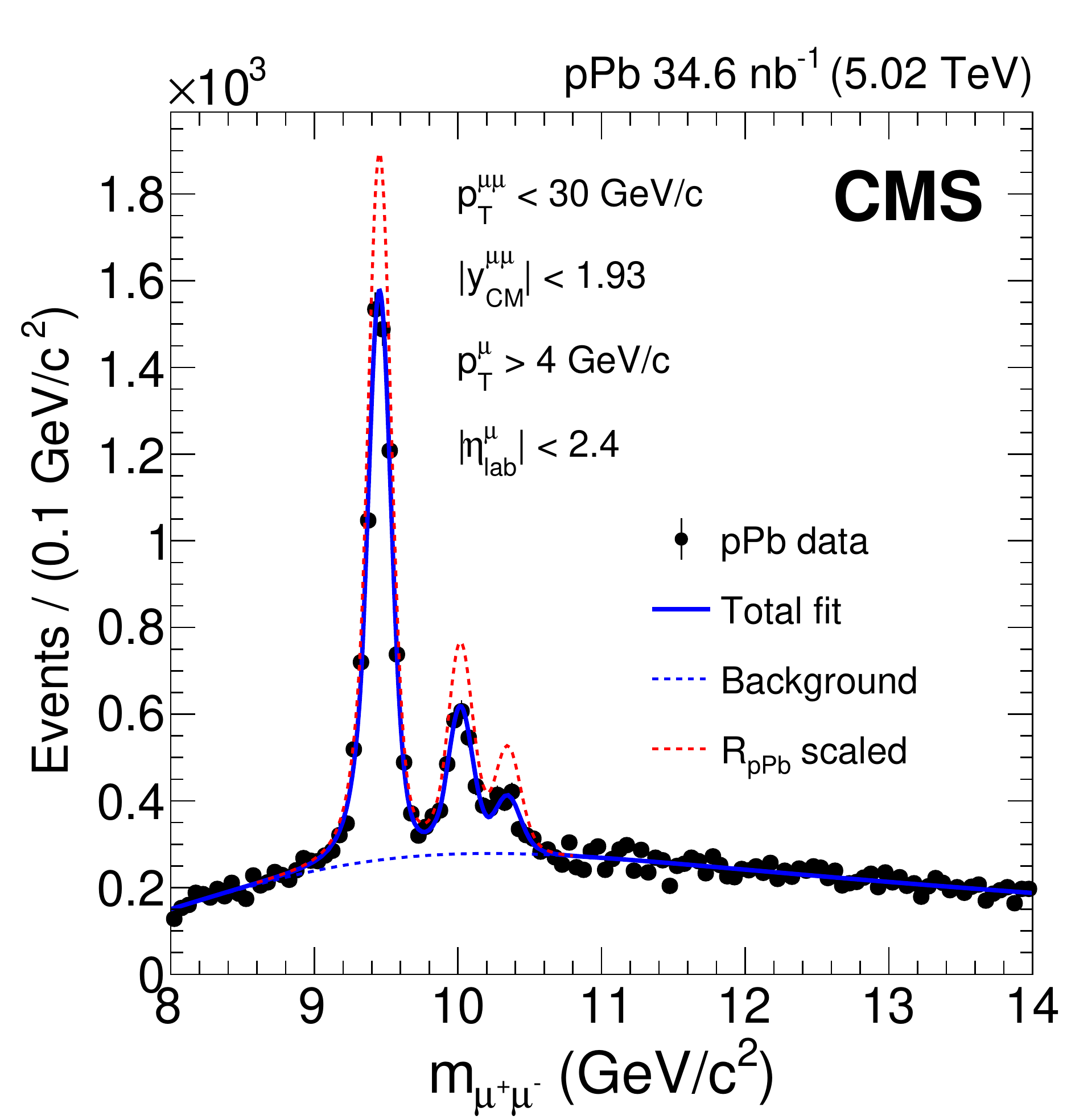}
\caption{
Measured dimuon invariant mass distributions (closed circles) for \pp (\cmsLeft) and \pPb (\cmsRight) collisions. The total unbinned maximum-likelihood fits to the data are shown as solid blue lines, with the background component indicated by dashed blue lines. The individual \PgUa, \PgUb, and \PgUc signal shapes in \pp are depicted as dashed orange lines in the \cmsLeft panel. The dashed red line in the \cmsRight panel is obtained by scaling the \PgUa, \PgUb, and \PgUc signal shapes in \pPb (solid blue line) under the assumption that \RpPb is unity.
} 
\label{fig:InvMass}
\end{centering}
\end{figure}

Figure~\ref{fig:InvMass} shows the invariant mass distributions of opposite-sign muon pairs for \pp (\cmsLeft) and \pPb (\cmsRight) collisions. The dimuon data are integrated in the dimuon range $\ptmumu<30\GeVc$ and $\abs{\rapmumu}<1.93$. The yields of the \PgU states, uncorrected for detector acceptance and efficiency, are obtained via unbinned maximum-likelihood fits to the invariant mass spectra, shown as solid blue lines. 
A dashed red line is used in Fig.~\ref{fig:InvMass} (\cmsRight) to depict the expected \PgUa, \PgUb, and \PgUc yields under the $\RpPb=1$ hypothesis, obtained by scaling the signal shape of each state by the inverse of its finally measured \RpPb value (including the ratio of the efficiencies corresponding to \pp and \pPb collisions).
This comparison illustrates that the \YnS yields are suppressed in \pPb relative to \pp collisions in the integrated kinematic region. We bin the data in the dimuon kinematic variables \ptmumu and \rapmumu, as well as in event activity variables which we discuss below. 

Quarkonium peaks can be modeled by a Crystal Ball (CB) function~\cite{SLAC-R-236}, whose low-mass power-law tail accounts for dimuons that undergo bremsstrahlung radiation in the detector material as well as final-state radiation. We model the shape of each \PgU state with a sum of two CB functions.
A parameter representing the relative CB widths is left free in the fit, to accommodate muons with different momentum resolutions (depending on their \etalabmu).
The relative contributions of the two CBs are also allowed to vary, both in the kinematic and event activity variables.

To eliminate unnecessary degrees-of-freedom in the fits, the relative widths and relative contributions of the two CB functions are constrained to be the same for all three \PgU states, consistent with fits to simulated samples.
Furthermore, parameters governing the shape of the radiative tail are constrained to be the same for all six CB functions in the fit.
The mass parameter of the \PgUa is left free to account for possible systematic shifts in the momentum scale of the reconstructed tracks. The final value of this parameter is consistent between fits to \pp and \pPb data.
Since any changes in the momentum scale should affect all measured \YnS similarly, we constrain the masses of the excited states such that their ratio matches the Particle Data Group (PDG) world-average values~\cite{PDG2018} as follows: $(m(n\mathrm{S})/m\mathrm{(1S)})_{\text{fit}} = (m(n\mathrm{S})/m\mathrm{(1S)})_{\mathrm{PDG}}$.
Similarly, the CB widths are also scaled by the ratio of the PDG mass values.

The parameters governing the tail shapes and ratio of CB widths are found to be correlated across kinematic bins.
Rather than allowing the parameters to be completely unconstrained, we instead allow them to vary around their mean values within an interval estimated from a set of preliminary fits.
The deviation of each parameter is translated into a Gaussian probability that is multiplied with the fit likelihood.
The width of each Gaussian function is set to the RMS value of the corresponding parameter in the preliminary fits.
In the case of \pPb collisions, the central value of the parameter determining the relative contributions of the two CB functions is constrained in this manner as well.

As a result of the large number of free parameters in the preliminary fits, it is possible for parameters to converge to different values in repeated fits.
By fitting the data across all the analysis bins, we find certain parameter values to be normally distributed.
Each normally distributed parameter is first restricted to its mean value across the preliminary fits, in order to enable the rest of the parameters to converge consistently across the bins. 
We take an iterative approach to this constraining technique to avoid biasing the final parameter values. 
The mean values of parameters from preliminary fits are obtained separately in different rapidity regions to allow for differences in \PgU meson reconstruction resolution
in the barrel and end-cap regions of the detector, where muons pass through different amounts of material and are detected using different technologies.

The background is modeled with a shifted and scaled error function multiplied by an exponential. 
The exponential function models the dominant combinatorial background, which falls with increasing dimuon invariant mass according to a statistical phase space factor.
The use of an error function is motivated by the effect of the $\ptmu > 4\GeVc$ selection applied to single muons, which produces 
a hump-like feature in the combinatorial background
at low invariant masses and at low dimuon \ptmumu. For dimuon $\ptmumu > 6\GeVc$, this feature 
moves to lower invariant masses outside the fit region
and we model the background solely with an exponential function. 

\subsection{Acceptance and efficiency corrections}
The \YnS yields that are extracted using fits to the invariant mass spectra are corrected to account for geometric limitations of the detector and inefficiencies of the online and offline selection algorithms. 
The dedicated MC simulations of \YnS decays
are used to determine the acceptance, which is the fraction of generated \PgU mesons in a given kinematic region that decay to muons satisfying the kinematic requirements applied in this analysis.

The efficiency of dimuon reconstruction, event triggering, and muon identification are studied using dedicated MC simulations of \YnS decays,
after they have undergone full detector response simulation. 
The dimuon efficiency is determined as the fraction of generated \PgU mesons in simulation that are identified as such, having satisfied all the same conditions that are required of muon pairs in collision data.
Since pure \PYTHIA-based MC samples are used for \pPb collisions, we verify that the efficiency correction does not exhibit any dependence on multiplicity.
This was also found in the related study of charmonium states reconstructed via muon pairs in \pPb collisions~\cite{HIN-14-009}.

Additional corrections are estimated to compensate for possible discrepancies between simulation and data efficiencies.
To estimate such discrepancies, muon triggering, track reconstruction, and identification efficiencies are measured using single muons from prompt \JPsi meson decays in both simulation and data, as described in Ref.~\cite{Chatrchyan:2012xi}. The ratios of the single-muon detection efficiencies between \JPsi data and simulation are estimated. These ratios differ significantly from unity in the case of muon triggering and identification in \pPb collisions and muon triggering and track reconstruction in \pp collisions.
These ratios are used to correct the simulation-based efficiencies.
For the bulk of muons in this analysis this correction to the efficiencies is small (${\sim}1\%$, but for some regions of phase space it can grow to at most 10\%).

\subsection{Systematic uncertainties}
Typical ranges of total and individual sources of systematic uncertainties in \RpPb and \RFB for all three \YnS states are tabulated in Table~\ref{tab:systematics}.
Two important sources of systematic uncertainty in the \YnS yields originate from an incomplete knowledge of the signal and background shapes. 
The signal shape systematic uncertainty is estimated using an alternative fit model of the signal, consisting of a single CB function in combination with a Gaussian function. 
This alternative fit model provides a comparable goodness-of-fit to the nominal one. 
For the background uncertainty estimation, a similar method of recalculating yields using an alternative fit model for the background is used. 
Because the background shape evolves with \ptmumu, the model was varied in different kinematic regions. 
In higher \ptmumu\ regions, a power law is used as the alternative background fit model.
In lower \ptmumu\ regions,
the background model is constructed from a linear combination of four invariant-mass fits to four \ptmumu\ subintervals of a MC simulation of dimuon decays.

When estimating systematic uncertainties in the \YnS yields using nominal and alternative models for signal and background distributions, we employ a method that helps to reduce the contribution of statistical fluctuations.
We perform pseudo-experiments where we generate a set of invariant mass distributions by MC sampling the shape fitted to the dimuon invariant mass spectra in each analysis bin.
Each generated invariant-mass distribution is fitted separately with the nominal and alternative signal models, using the nominal background model in both cases. 
The systematic uncertainty is evaluated as the mean of absolute values of relative differences between yields extracted using the nominal and alternative models.
Similarly, additional pseudo-experiments are performed to estimate the uncertainty associated to the choice of the background model.

The procedure for constraining the parameters of the signal model introduces another source of systematic uncertainty. 
In order to estimate the systematic uncertainty on the yields we perform a set of preliminary fits to the dedicated \YnS MC simulations, in which the parameter phase space is iteratively reduced in the same way as for data.
For each parameter, we use the mean value obtained from the last iteration of MC fits as an alternative value for the mean of the Gaussian probability function used to constrain the parameters when fitting to data.
In the case of \pPb, the value of the parameter determining the relative contributions of the two CB functions in the free-parameter fit to MC is used as its alternative value. 
We compare the yields from the nominal fits with those extracted using the alternative mean of the constraining Gaussians and calculate the deviations of the yields for each constrained parameter. The largest of these deviations is assigned as the systematic uncertainty.

Systematic uncertainties in the acceptance corrections are estimated by varying the parameters of the fit used to weight MC \ptmumu\ spectra within uncertainties, and recording the largest produced deviation.
For the \RpPb measurement, the Y states are assumed to be unpolarized in both \pp and \pPb collisions. We also assume that if the production mechanism of \YnS leads to a different polarization, then it remains the same for both \pp and \pPb collisions. As a consequence, the acceptance ratio under a different polarization scenario will remain equal to unity. Therefore the uncertainty in the polarization would not affect the nuclear modification factor.

Systematic uncertainties in efficiency corrections are estimated by combining two sources in quadrature. The first is the uncertainty in weighting MC \ptmumu\ spectra, which is estimated by determining the efficiency using the MC samples with and without weighting.
The second source of uncertainty arises from the data-to-simulation ratio of single-muon detection efficiencies that are used to correct the purely simulation-based dimuon efficiencies.
The systematic uncertainty in the efficiency of each stage of muon detection (triggering, tracking, muon identification) is studied by varying the selection criterion for that stage, while the statistical uncertainty is determined by repeating such variations one hundred times and estimating the standard deviation.

The total systematic uncertainty from uncorrelated sources is obtained by combining the uncertainties in quadrature 
in the signal and background extractions, as well as in the acceptance and efficiency corrections.
The combined systematic uncertainty in the results increases slightly with increasing \abs{\rapUps} and with decreasing \ptUps. 
Because of the asymmetry of \pPb collisions, the most forward \rapUps bins, which are at the edge of the detector, have larger systematic uncertainty than the most backward \rapUps bins because the latter are closer to $\ylabUps = 0$. 
The total systematic uncertainty also increases with increasing event activity for integrated \ptUps\ and \rapUps. 

\begin{table}[hbtp]
\centering
\topcaption{Ranges of typical systematic uncertainties in \RpPb and \RFB for \YnS. For acceptance and efficiency the range quoted 
covers the efficiency of all three \PgU states. The uncertainties in luminosity are global uncertainties that apply to all three \PgU states.
The luminosity uncertainty cancels in the calculation of the \RFB.}
\begin{tabular}{cccc}
\hline
Source & \PgUa & \PgUb & \PgUc \\
\hline
 &  & \RpPb &  \\
Background & 0.3--12\% & 1--6\% & 2--8\% \\
Signal & 1--8\% & 2--10\% & 3--9\% \\
Acceptance & & ${\lesssim}1\%$ & \\
Efficiency &  & 4--6\% &  \\
Luminosity (\pPb) &  & 3.5\% &  \\
Luminosity (\pp) &  & 2.3\% &  \\[\cmsTabSkip]
&  & \RFB &  \\
Background & 2--4\% & 4--7\% & 5--7\% \\
Signal & 2--3\% & 2--5\% & 5--6\% \\
Acceptance &  & ${\lesssim}1\%$ &  \\
Efficiency &  & ${\approx}2\%$ &  \\
\hline
\end{tabular}
\label{tab:systematics}
\end{table}

\section{Results}
\begin{figure*}[tbhp!]
\centering
\includegraphics[width=0.48\textwidth]{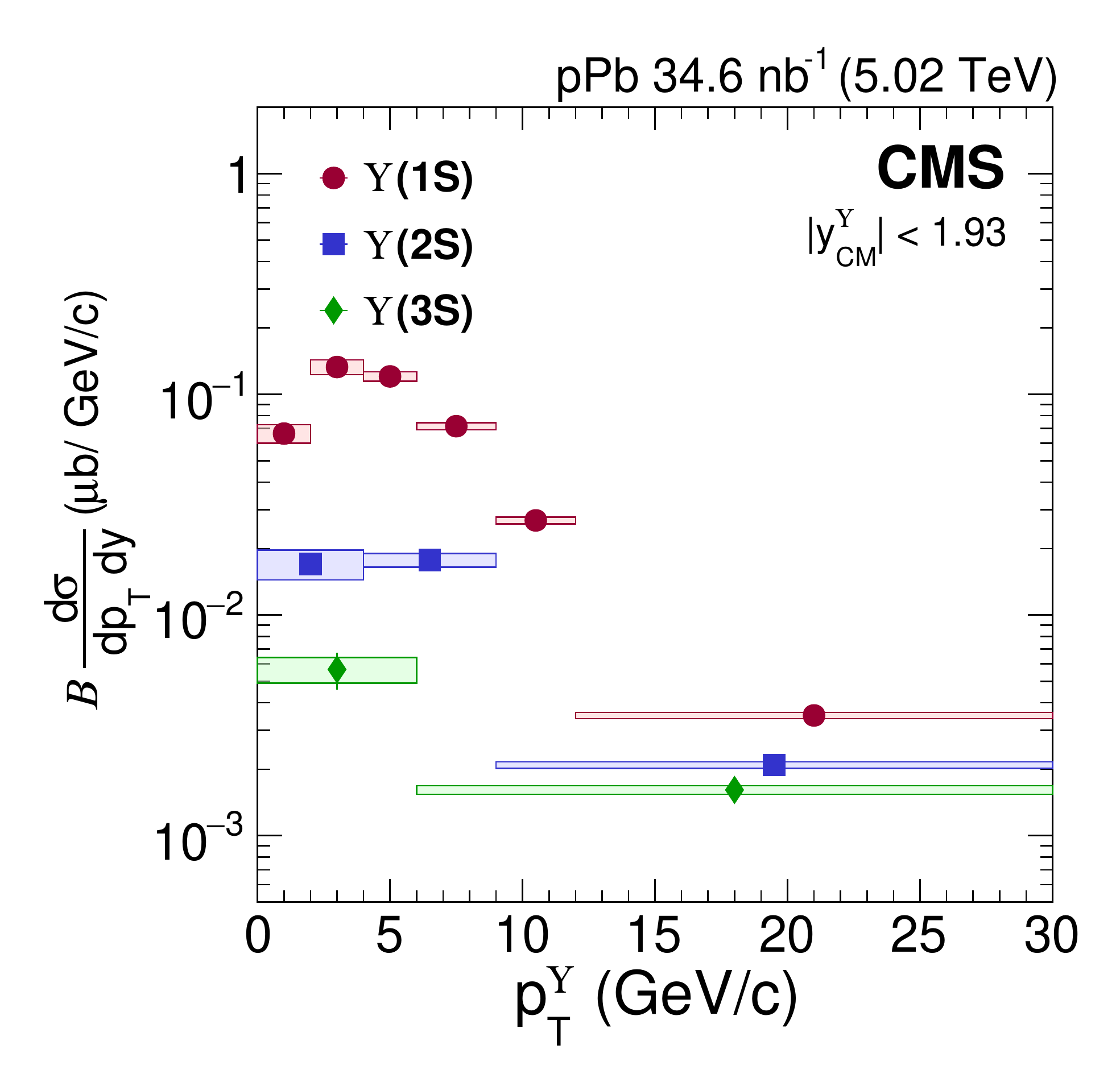} 
\includegraphics[width=0.48\textwidth]{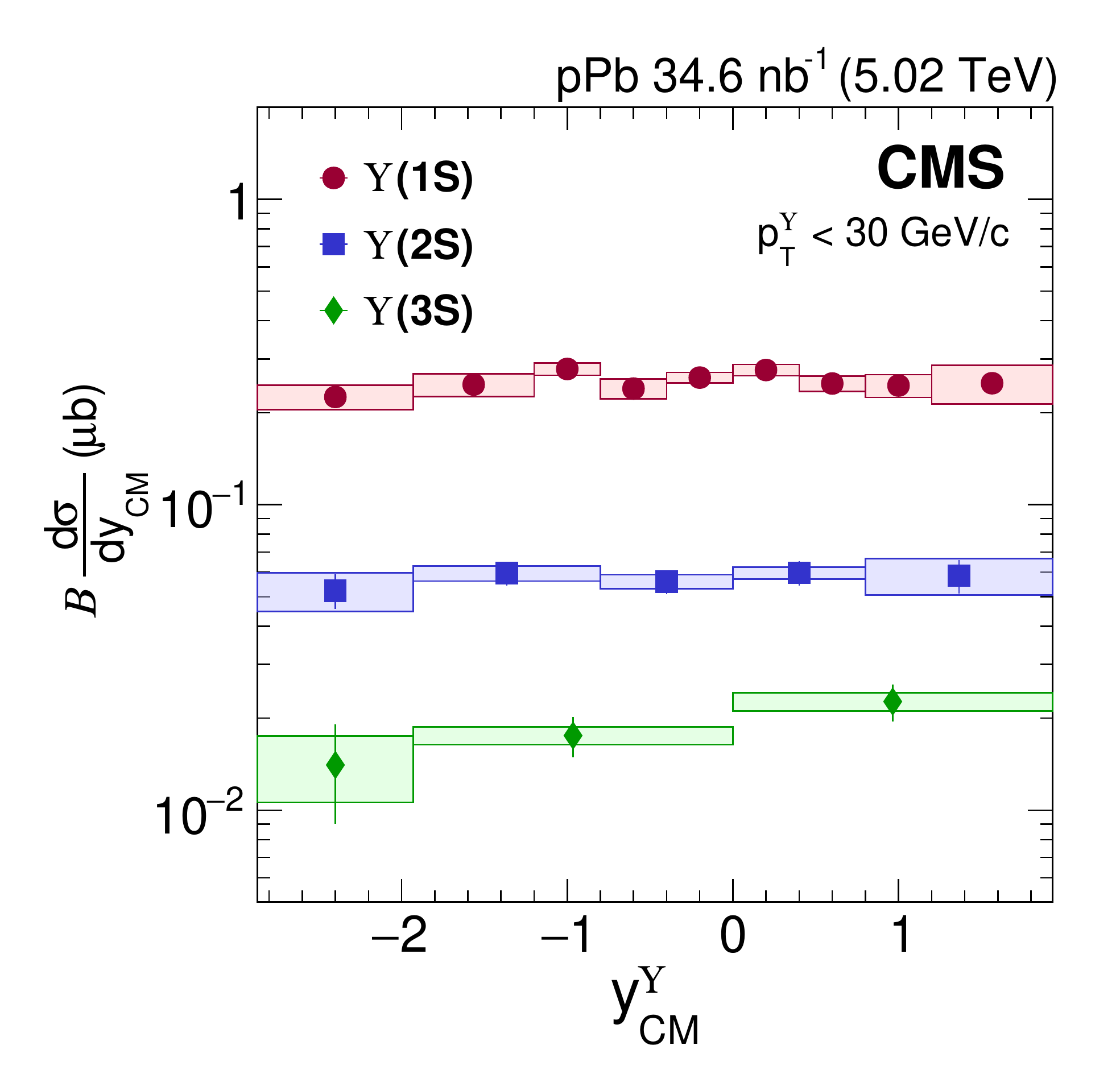}
\includegraphics[width=0.48\textwidth]{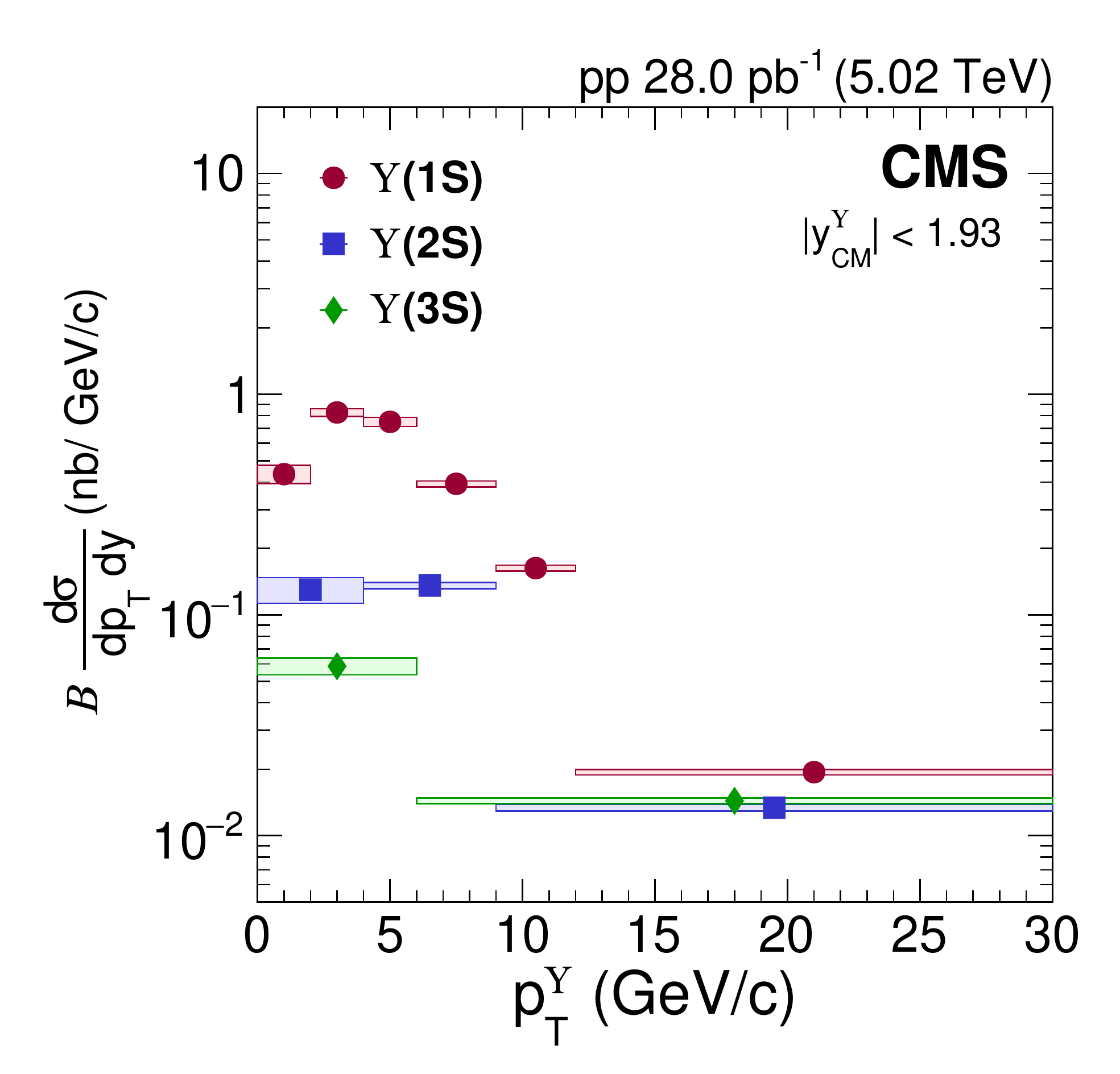} 
\includegraphics[width=0.48\textwidth]{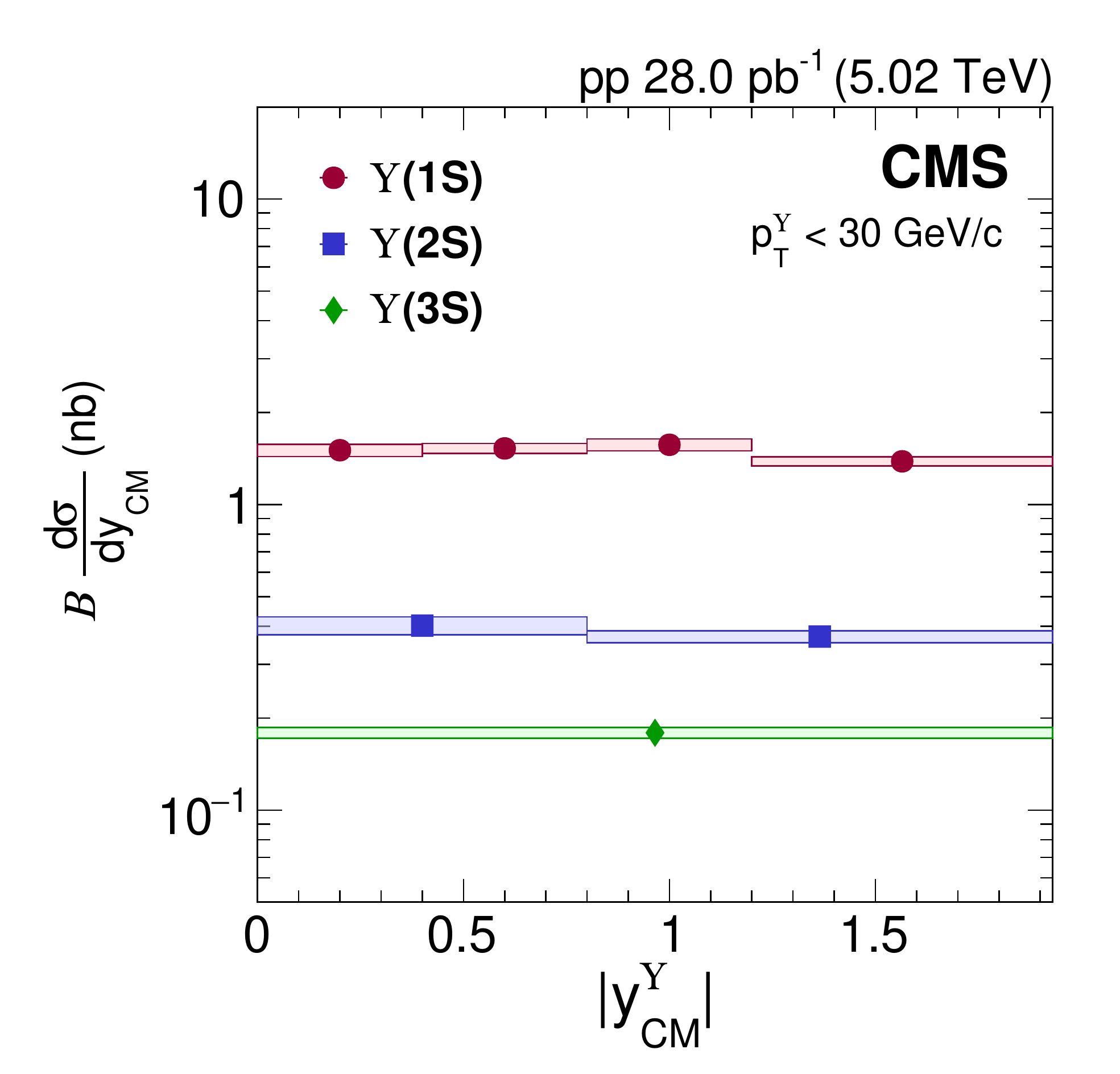}
\caption{Cross section times dimuon branching fraction of \PgUa (red circles), \PgUb (blue squares), and \PgUc (green diamonds) as functions of \ptUps\ (left) and \rapUps (right) in \pPb (upper row) and \pp (lower row) collisions. For \pPb collisions, the p-going side corresponds to $\rapUps>0$. Because pp collisions are symmetric in the center-of-mass frame, the absolute value of rapidity $\abs{\rapUps}$ is used in the lower right panel. 
Vertical bars represent statistical and fit uncertainties and filled boxes represent systematic uncertainties.
A 3.5 (2.3)\% global uncertainty in determining the integrated luminosity of \pPb (\pp) collisions, applicable to all points, is not included in the point-by-point uncertainties.
}
\label{fig:XS_pPb_PP}
\end{figure*}

The product of the branching fraction of \YnS to muon pairs, $\mathcal{B}(\YnS\to\muplusmuminus)$, and the double-differential production cross section, $\rd^2\sigma/\rd\ptUps\rd\rapUps$, is obtained as
\begin{equation}
\mathcal{B}\bigl(\YnS\to\muplusmuminus\bigr)\frac{\rd^2\sigma}{\rd \ptUps \rd \rapUps} = \frac{N^{\YnS}_{\text{Fit}}/(a \varepsilon)}{\Lumi_{\text{int}}\Delta \ptmumu \Delta \rapmumu},
\label{eqn:crosssection}
\end{equation}
where 
$N^{\YnS}_{\text{Fit}}$ is the yield of \YnS mesons extracted from the fit in a given (\ptmumu, \rapmumu) bin, 
$a$ is the dimuon acceptance correction, $\varepsilon$ is the dimuon efficiency correction,
and $\Lumi_{\text{int}}$ is the integrated luminosity. 
Figure~\ref{fig:XS_pPb_PP} (upper row) shows the cross sections of \PgUa, \PgUb, and \PgUc 
in \pPb collisions as functions of \ptUps\ (left) and \rapUps (right).
The error bars on the points are those from the fits to obtain the yields, which take into account the Poisson statistical uncertainties in the invariant mass distribution and the uncertainties associated with correlations between the parameters used in the probability density functions to fit the data. The filled boxes represent the systematic uncertainties, as discussed in the previous section.
When investigating the \pPb cross section dependence on \ptUps\ and when determining the \RpPb and \RFB, we restrict the CM rapidity range to the symmetric region $\abs{\rapmumu}<1.93$ where data for both \pp and \pPb collisions are available.
The \YnS cross sections in \pp collisions are also determined for $\abs{\rapUps}<1.93$ as functions of the kinematic variables. These are shown in Fig.~\ref{fig:XS_pPb_PP} (lower row).

\begin{figure}[tbh!]
\centering
\includegraphics[width=0.48\textwidth]{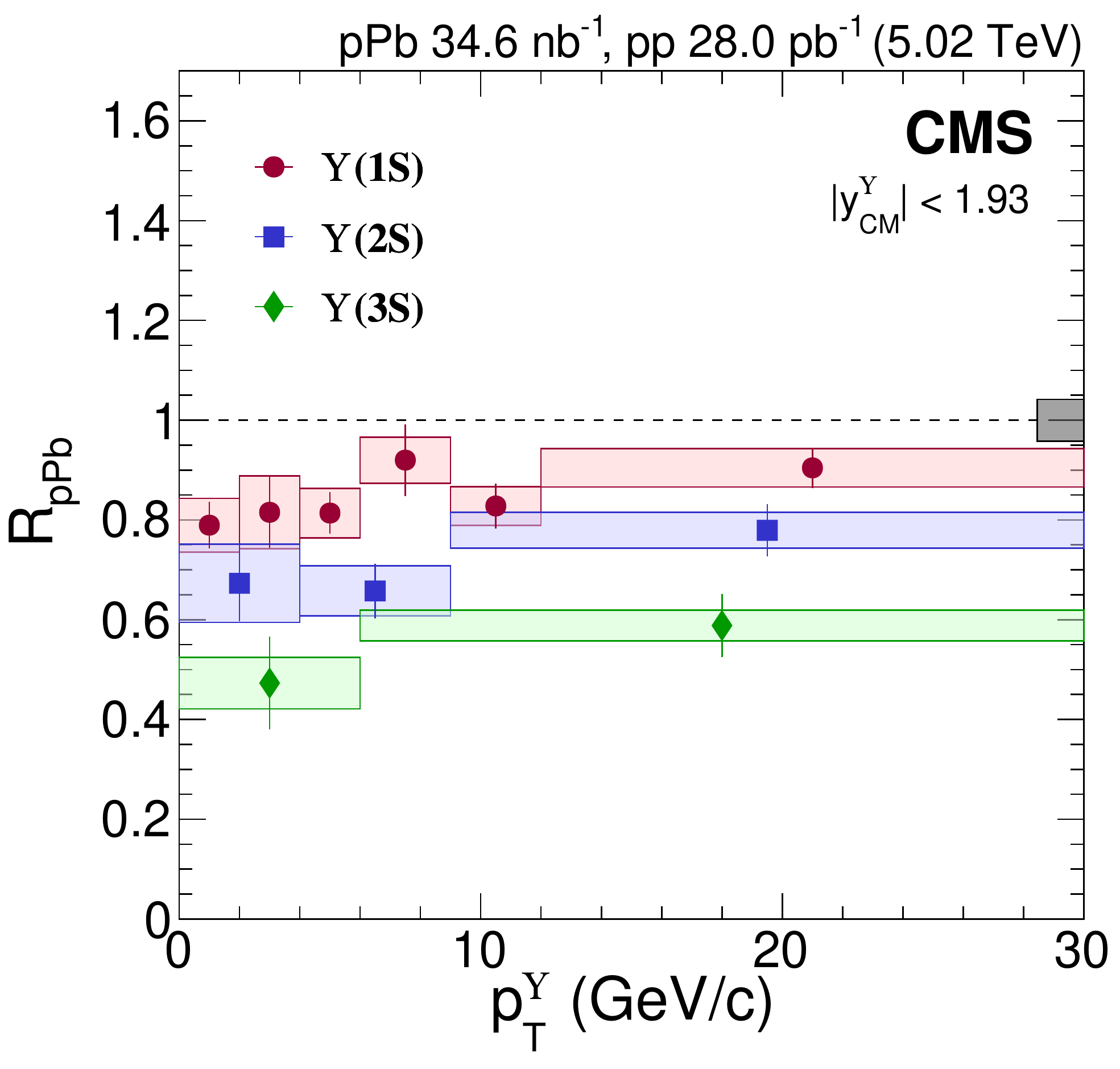}
\includegraphics[width=0.48\textwidth]{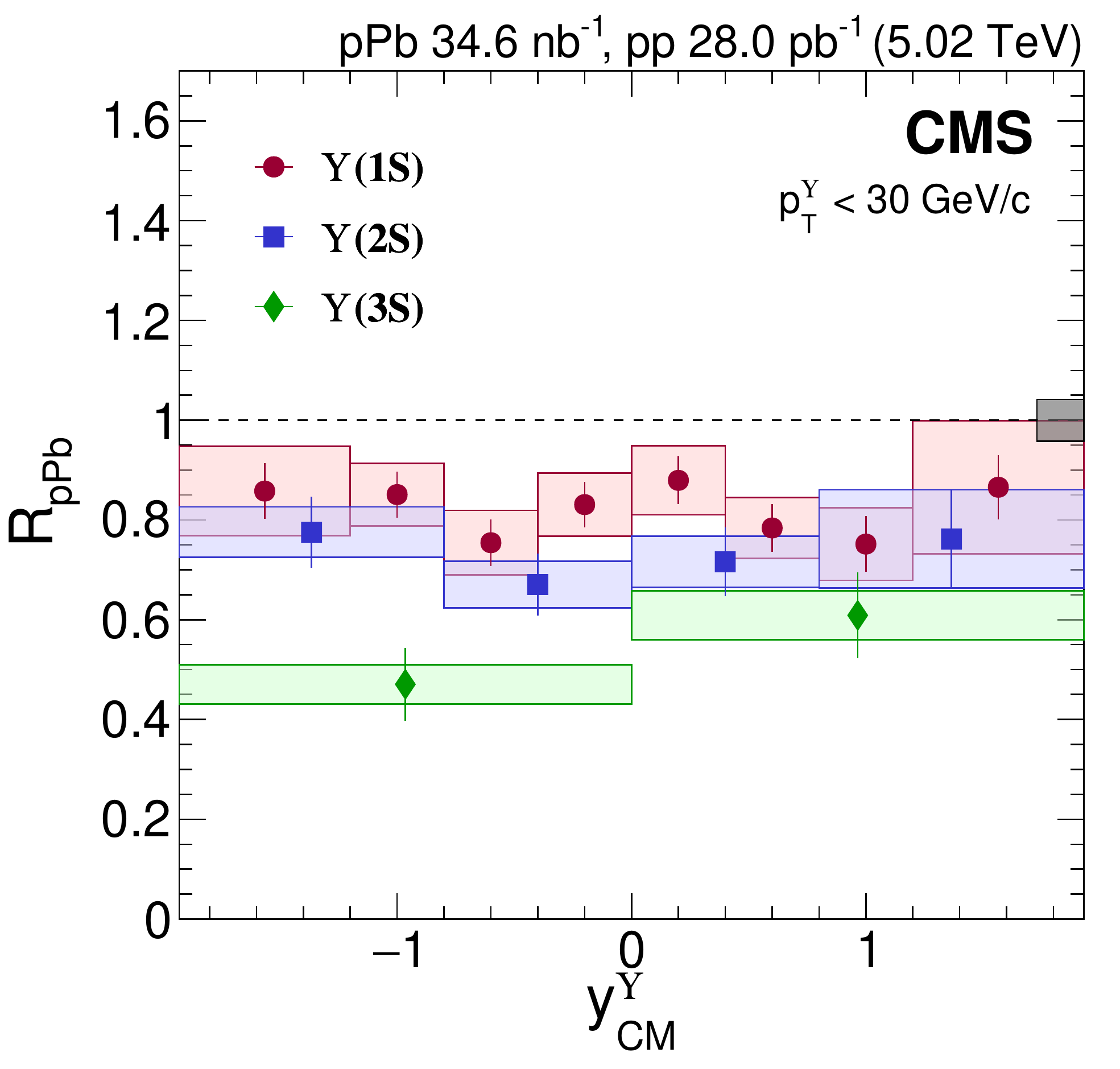}
\caption{\RpPb of \PgUa (red circles), \PgUb (blue squares), and \PgUc (green diamonds) as functions of \ptUps\ (\cmsLeft) and \rapUps (\cmsRight), where the \cmsRight panel is integrated over $\ptUps<30\GeVc$. Vertical bars on the points represent statistical and fit uncertainties and filled boxes represent systematic uncertainties. The gray box around the line at unity represents the global uncertainty due to luminosity normalization (4.2\%).
}
\label{fig:RpA1D}
\end{figure}

The quantity \RpPb is calculated as
\begin{equation}
\label{eqn:rpa}
\RpPb\bigl(\ptUps,\rapUps\bigr) = \frac{({\rd^2\sigma}/{\rd \ptUps \rd \rapUps})_{\pPb}}{A({\rd^2\sigma}/{\rd \ptUps \rd \rapUps})_{\pp}},
\end{equation}
where $A=208$ is the mass number of the Pb nucleus. Results for \RpPb are shown in Fig.~\ref{fig:RpA1D} as functions of \ptUps\ and \rapUps. 
We observe that all three \YnS states are suppressed in \pPb relative to \pp collisions throughout the kinematic region explored, suggesting modification by CNM effects in \pPb collisions. 
Similar to the \PbPb case~\cite{HIN-16-023}, the level of suppression for each \PgU state in \pPb collisions is consistent with a constant value in the kinematic region studied, although the level of suppression seen in \PbPb is much stronger.
The ATLAS Collaboration reported an increasing \RpPb with \ptUps\ for \PgUa~\cite{ATLASRpA} in a similar midrapidity region as in CMS. 
The CMS data is consistent with no dependence, but the overall \ptUps\ dependence of the \RpPb(\PgUa) in the two experiments is consistent within uncertainties. Moreover, our data shows no \rapUps dependence, which is consistent with the ATLAS result.

In the charmonium sector, the CMS Collaboration found hints of an ordered suppression pattern. The \RpPb of \Pgy\ was found to be smaller than that of \JPsi~\cite{HIN-14-009} in \pPb collisions at $\sqrtsNN = 5.02\TeV$ for backward rapidity and $\pt^{\scriptscriptstyle{\JPsi}} < 10\GeVc$~\cite{HIN-16-015}.
The results presented here suggest a similar ordered suppression of the \PgU states in the backward rapidity region as well as across the entire \ptUps\ region studied. The measured \RpPb(\PgUa) is systematically larger than that of \PgUb, which in turn is systematically larger than the \RpPb(\PgUc), suggesting different levels of modification to the three states by final-state effects in these regions. In the forward rapidity region, the measured \RpPb of the three states appear more mutually consistent.

We further compare the \rapUps dependence of the measured \RpPb to predictions from three CNM models: shadowing, energy loss, and comover interaction.
The shadowing calculations incorporate next-to-leading order nuclear modifications of the PDFs (nPDFs)~\cite{VogtComp}, according to EPS09~\cite{EPS09}.
Predictions using coherent energy loss~\cite{ArleoComp} are made with and without using EPS09.
Since they affect the quarkonium system before hadronization, both shadowing and energy loss are initial-state effects. 
Finally, predictions using the comover interaction model (CIM)~\cite{FerreiroComp} are provided with two different leading-order nPDF calculations: EPS09 and nCTEQ15~\cite{nCTEQ15}. 
Since the comovers in the CIM interact with the quarkonium system after hadronization, it is deemed to be a final-state CNM effect.

\begin{figure}[tbh!]
\centering
\includegraphics[width=0.48\textwidth]{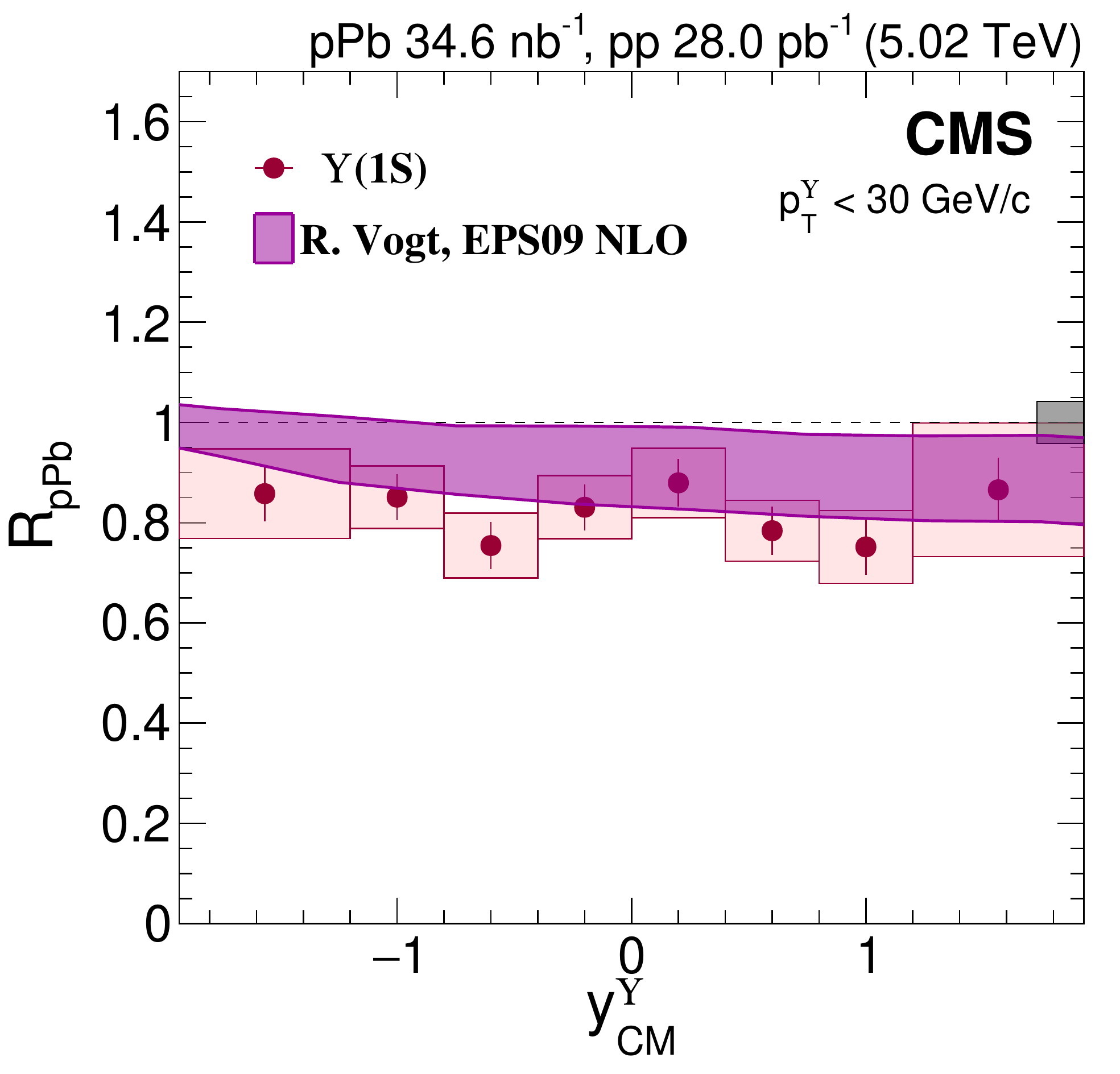}
\includegraphics[width=0.48\textwidth]{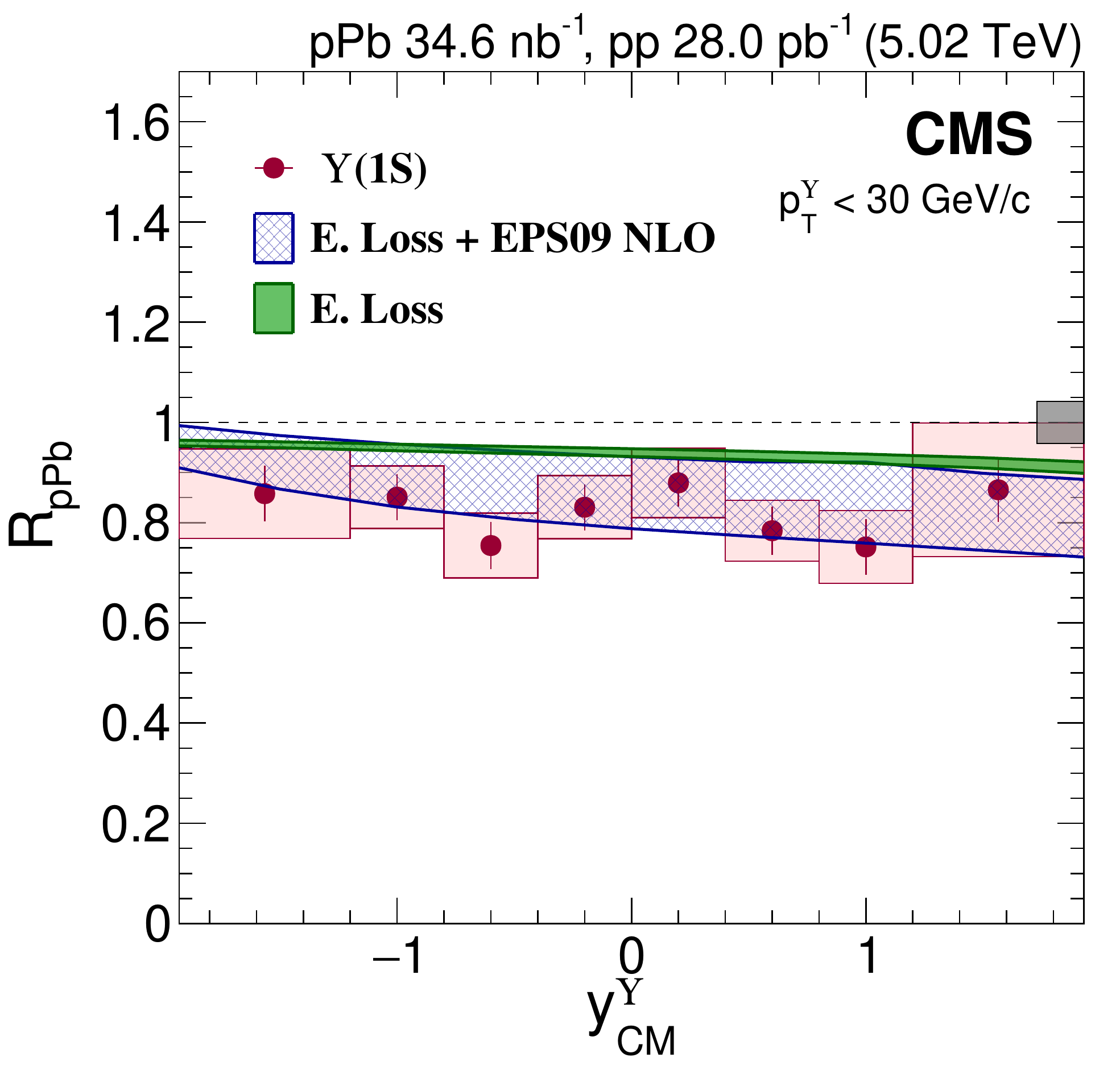}
\caption{\RpPb of \PgUa (red circles) versus \rapUps with initial-state model calculations: nPDF modification~\cite{VogtComp} (\cmsLeft) and energy loss (E. Loss) with and without shadowing corrections~\cite{ArleoComp} (\cmsRight). 
The uncertainty range for each model calculation is shown.
Vertical bars on the points represent statistical and fit uncertainties and filled boxes represent systematic uncertainties. The gray box around the line at unity represents the global uncertainty due to luminosity normalization (4.2\%).}
\label{fig:TheoryRpArapIni}
\end{figure}

In Fig.~\ref{fig:TheoryRpArapIni}, the measured \RpPb(\PgUa) is compared to predictions from shadowing~\cite{VogtComp} (\cmsLeft) and predictions using energy loss only and energy loss with shadowing~\cite{ArleoComp} (\cmsRight).
The uncertainty in the models comes from the nPDFs.
The combined energy loss with shadowing model is in better agreement with our data, but given the current uncertainties in theory and experiment, the models using only shadowing or energy loss cannot be ruled out. The shadowing model, which includes only initial-state effects, predicts equal modification of all bottomonium states and therefore is incompatible with the \PgUb and \PgUc data.
 
\begin{figure*}[tbhp!]
\centering
\includegraphics[width=0.48\textwidth]{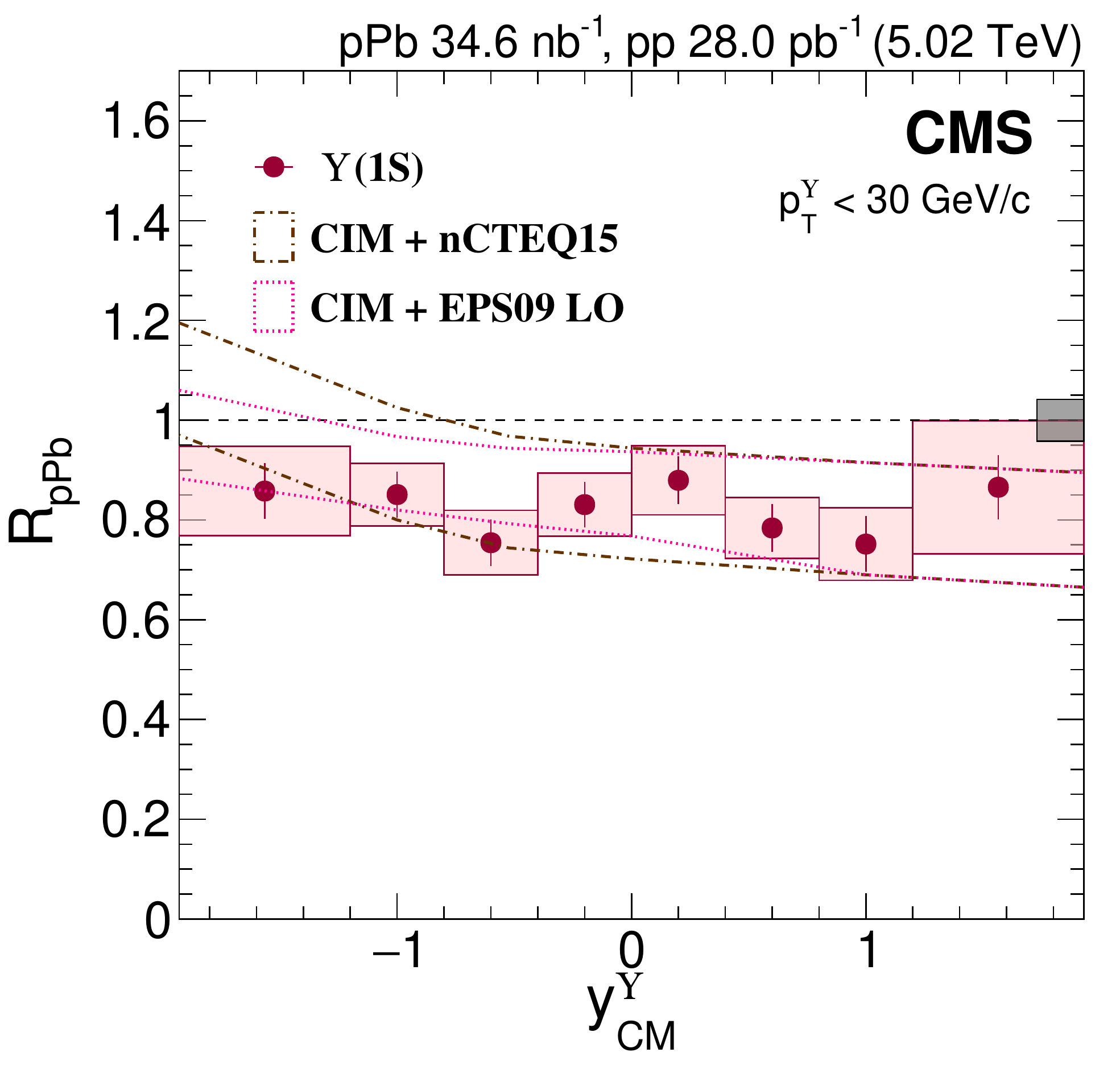}
\includegraphics[width=0.48\textwidth]{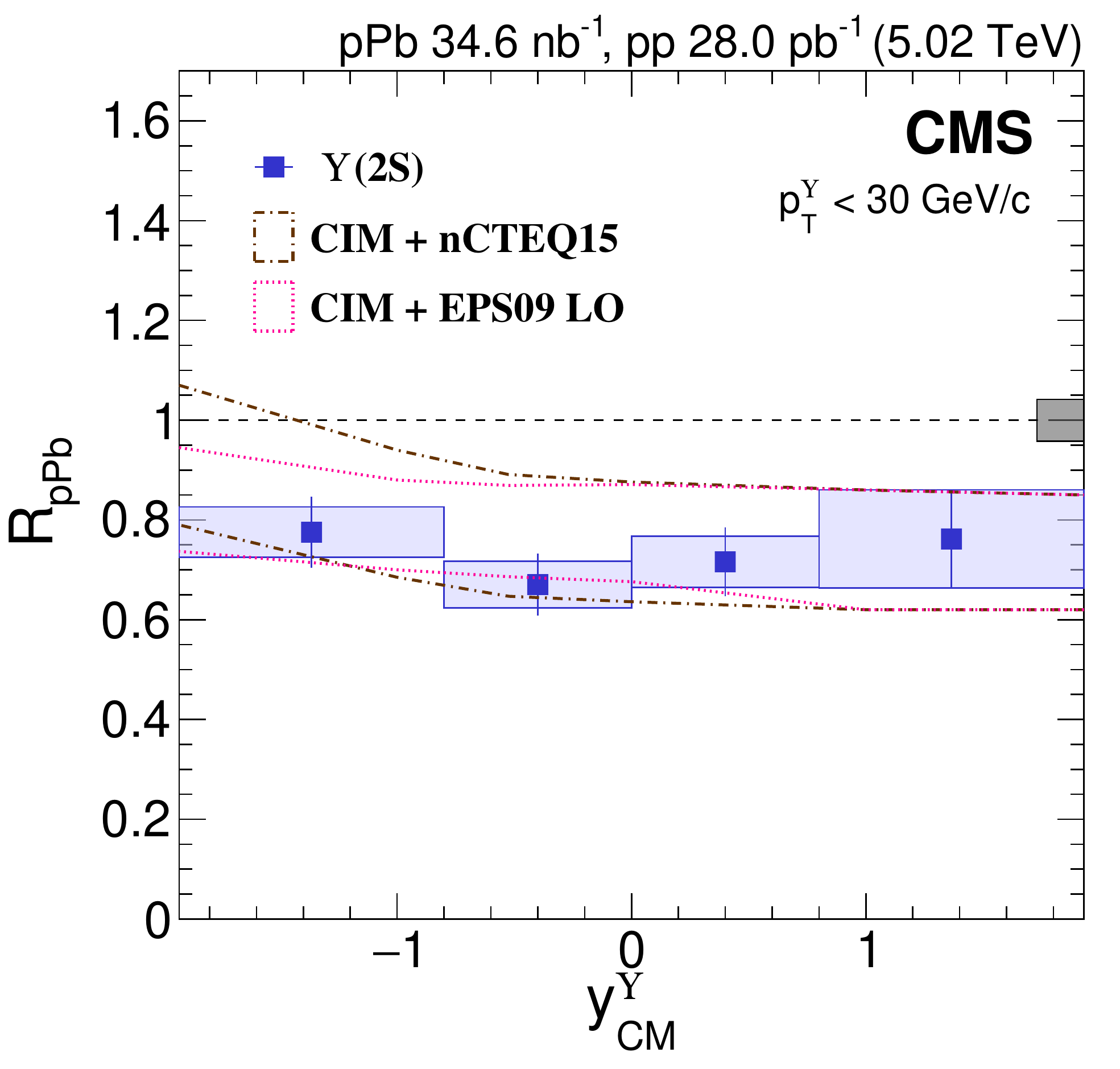}
\includegraphics[width=0.48\textwidth]{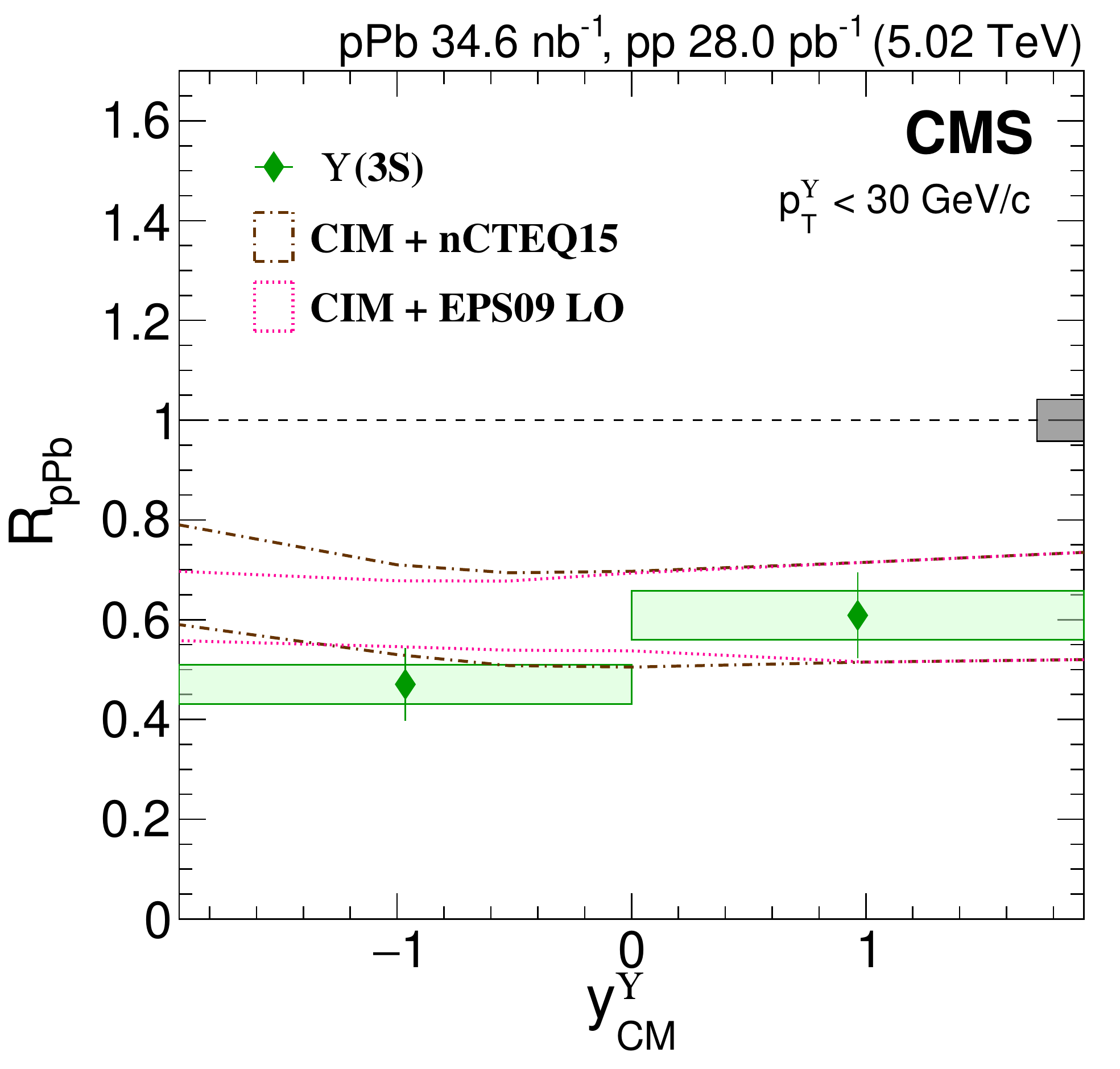}
\caption{\RpPb versus \rapUps with CIM predictions~\cite{FerreiroComp} with shadowing corrections using nCTEQ15 and EPS09 for \PgUa (upper left; red circles), \PgUb (upper right; blue squares) and \PgUc (lower; green diamonds). 
The uncertainty range for each model calculation is shown.
Vertical bars on the points represent statistical and fit uncertainties and filled boxes represent systematic uncertainties. The gray box around the line at unity represents the global uncertainty due to luminosity normalization (4.2\%).}
\label{fig:TheoryRpArapComover}
\end{figure*}

In contrast to shadowing and energy-loss models, the CIM predicts different degrees of modification for the \PgUa, \PgUb, and \PgUc states~\cite{FerreiroCharm15,FerreiroComp},
since higher excited states have a larger size and hence increased comover interactions.
In addition, comover modification of quarkonium states is expected to be stronger in regions where the comover densities are larger, such as in the nucleus-going direction in asymmetric proton-nucleus collisions and in regions of higher event activity.
Figure~\ref{fig:TheoryRpArapComover} shows comparisons of predicted \RpPb in the CIM~\cite{FerreiroComp}, including shadowing corrections from both nCTEQ15 and EPS09 nPDFs,
with the measured \RpPb for \PgUa (upper left), \PgUb (upper right), and \PgUc (lower).
The CIM \RpPb predictions show similar ordered suppression to that found in the data, an effect missing in models with only initial-state effects.

\begin{figure}[tbhp!]
\centering
\includegraphics[width=0.48\textwidth]{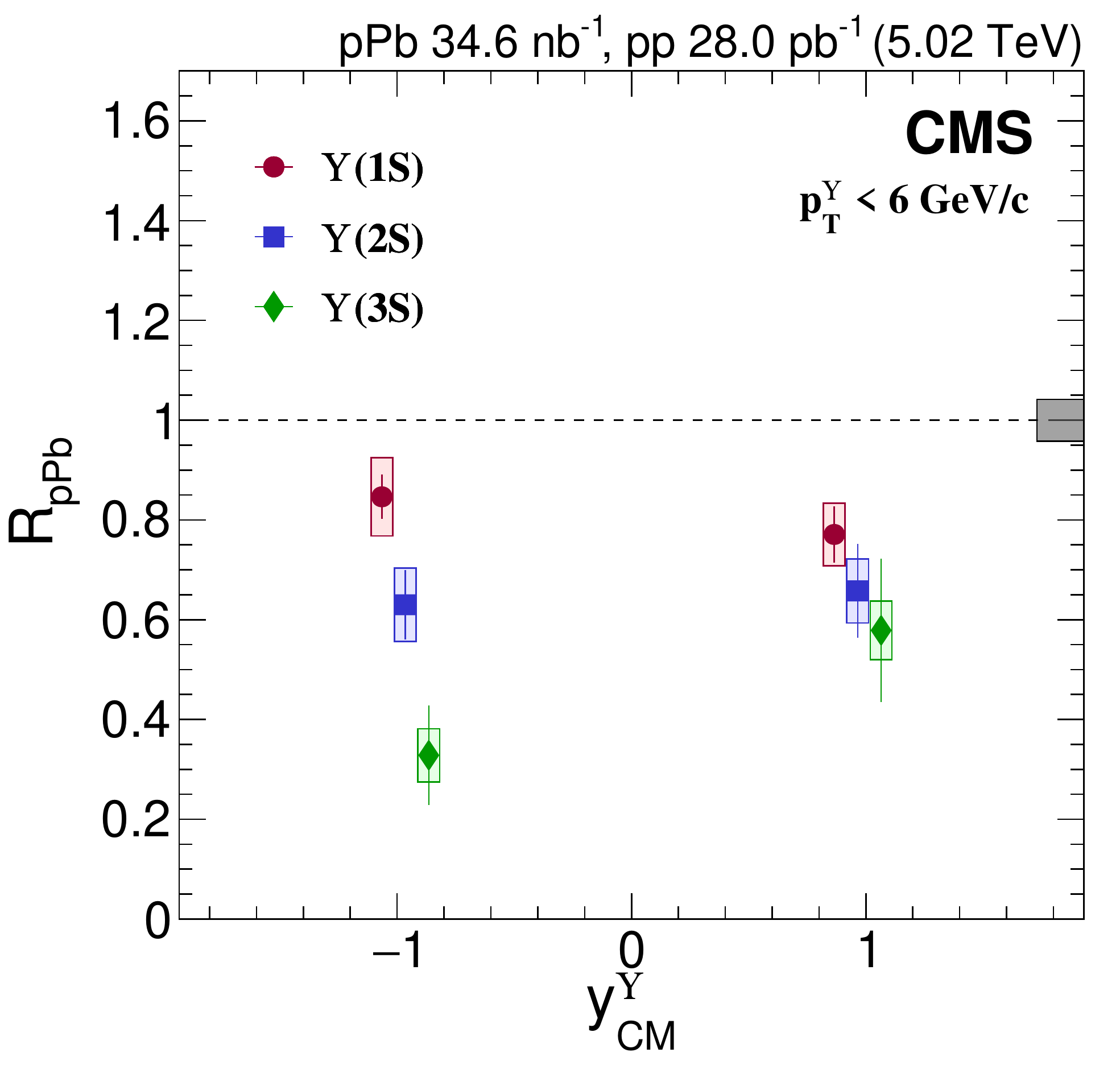} 
\includegraphics[width=0.48\textwidth]{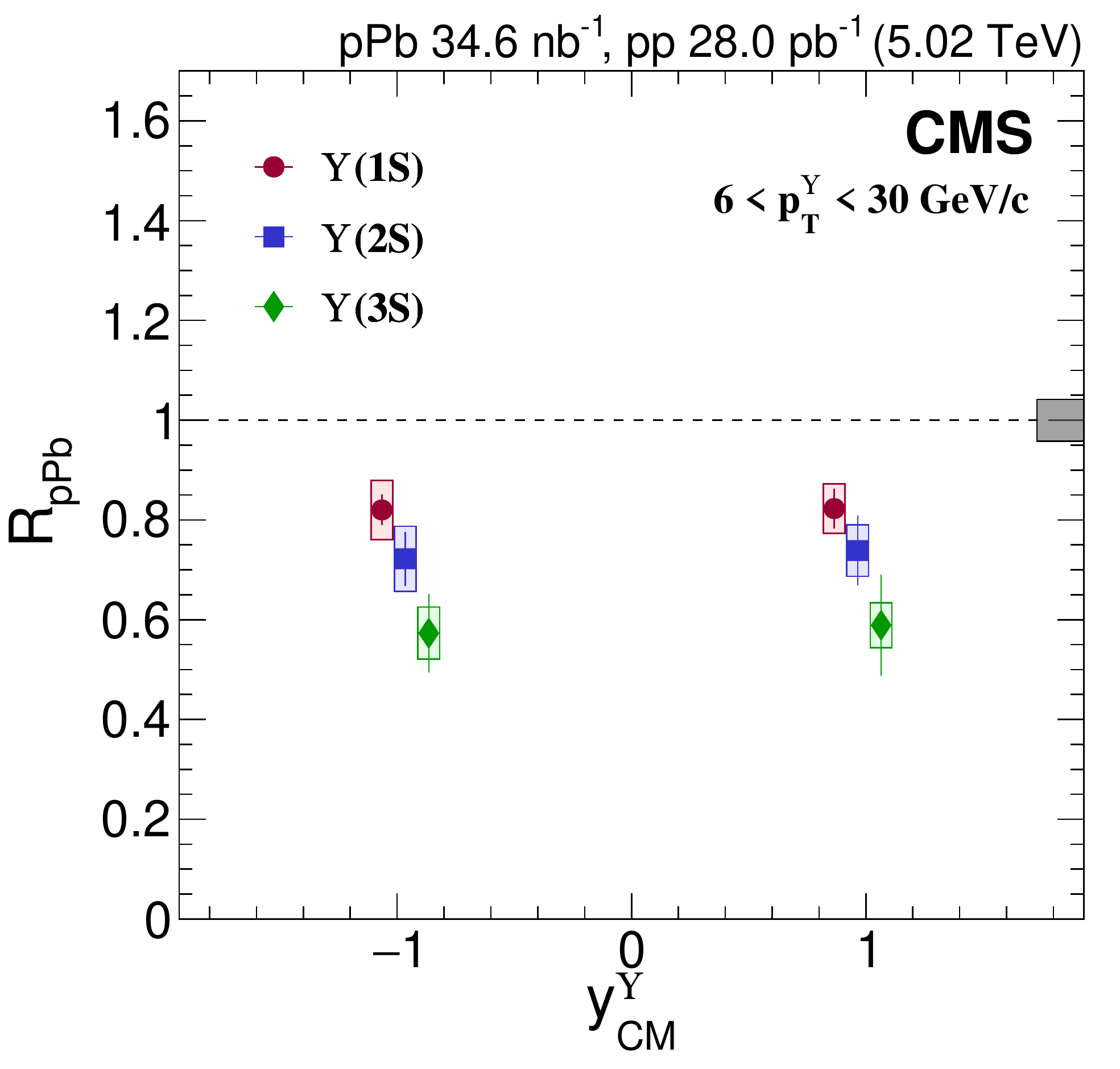}
\caption{\RpPb of \PgUa (red circles), \PgUb (blue squares), and \PgUc (green diamonds) at forward and backward rapidity for $0<\ptUps<6\GeVc$ (\cmsLeft) and $6<\ptUps<30\GeVc$ (\cmsRight). The points are shifted horizontally for better visibility. Vertical bars on the points represent statistical and fit uncertainties and filled boxes represent systematic uncertainties. The gray box around the line at unity represents the global uncertainty due to luminosity normalization (4.2\%).}
\label{fig:RpArap2D}
\end{figure}

By comparing the \RpPb(\YnS) in the forward (proton-going) and backward (lead-going) directions, we can investigate the dependence of bottomonium suppression on the amount of nuclear matter present.  
Figure~\ref{fig:RpArap2D} shows the \RpPb of \YnS states for $-1.93<\rapUps<0$ and $0<\rapUps<1.93$ in the low-\ptUps\ (\cmsLeft) and high-\ptUps\ (\cmsRight) regions. 
We find indication of greater differences between the suppression levels of low-\ptUps\ \YnS states in the lead-going versus the proton-going \rapUps directions. 
A similar observation was made by CMS in the charmonium sector~\cite{HIN-16-015}, where the modification levels of \Pgy\ and \JPsi\ with $\pt<10\GeVc$ were more separated in the backward region, whereas both states experienced similar modification in the forward region. 

\begin{figure}[tbhp!]
\centering
\includegraphics[width=0.48\textwidth]{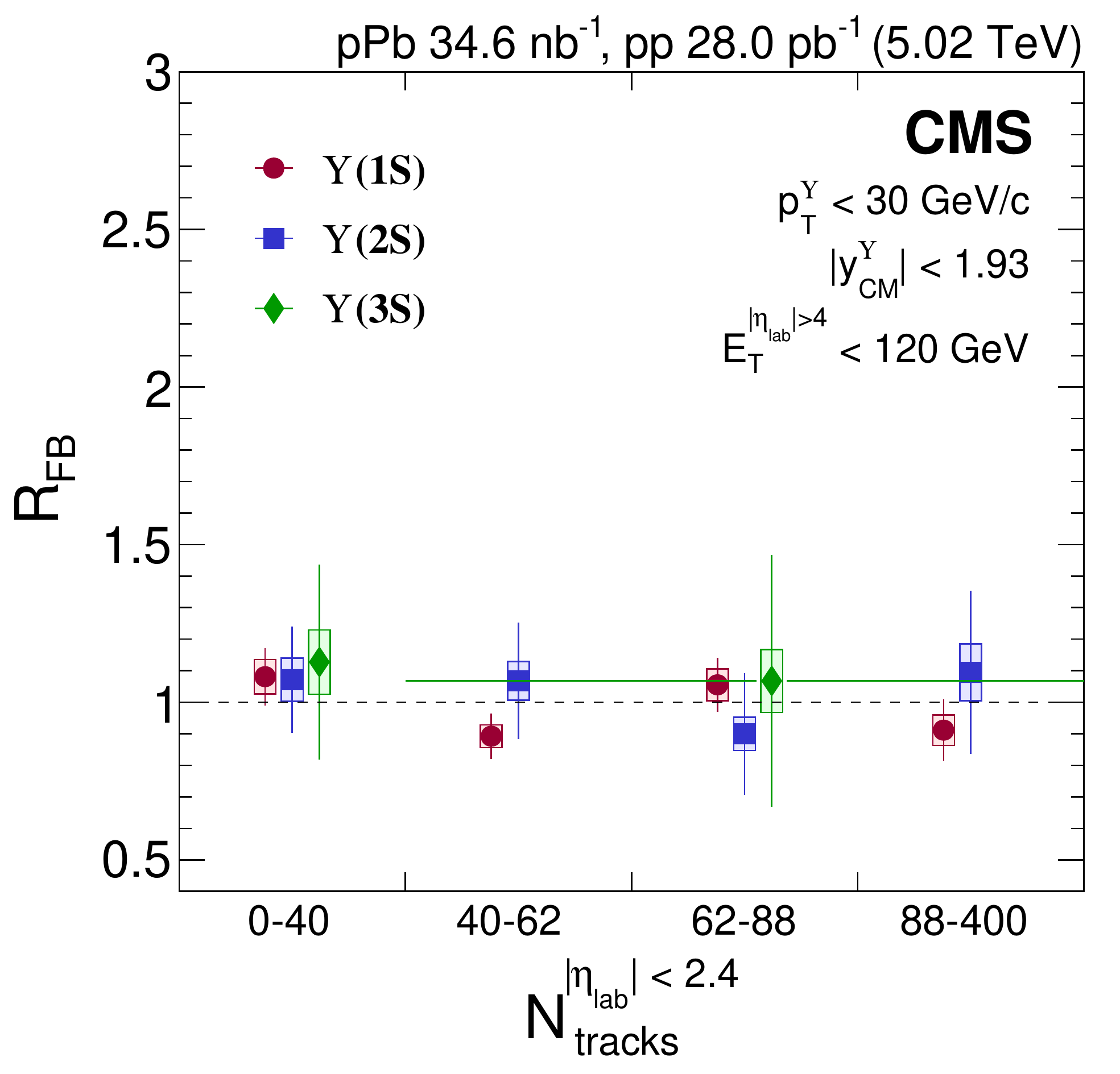}
\includegraphics[width=0.48\textwidth]{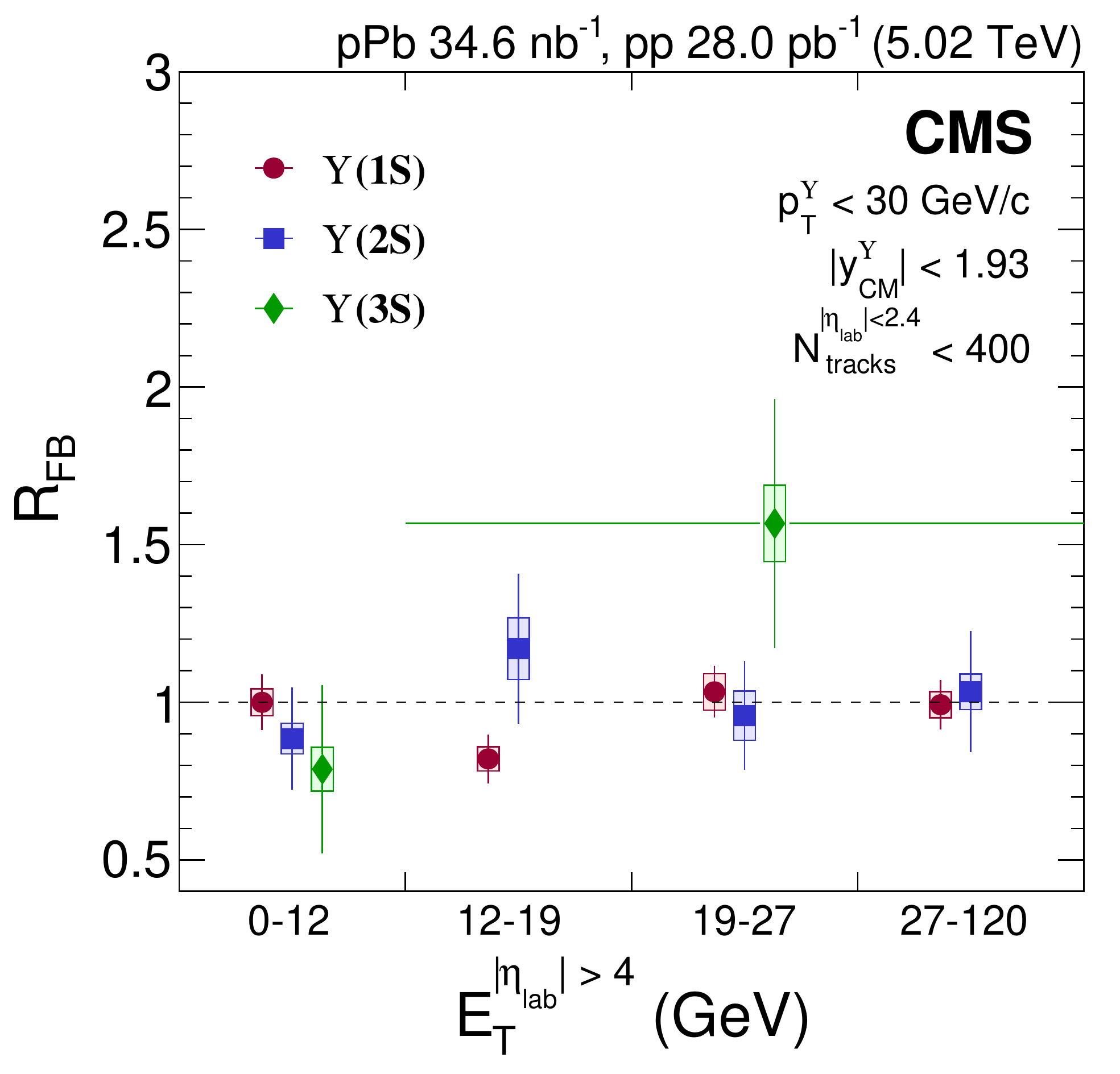}
\caption{\RFB versus \Ntracks at mid-pseudorapidity (\cmsLeft) and \vs \ET at forward/backward pseudorapidity (\cmsRight) of \PgUa (red circles), \PgUb (blue squares), and \PgUc (green diamonds) for $\ptUps<30\GeVc$ and $\abs{\rapUps}<1.93$. 
Vertical bars on the points represent statistical and fit uncertainties and filled boxes represent systematic uncertainties.
For \PgUc, a wide bin is used for high event activity, with the width indicated by a horizontal bar.
}
\label{fig:RFBvsHFNtracksInt}
\end{figure}

We study the forward-backward production ratio of \PgU mesons in \pPb collisions defined as follows,
\begin{equation}
\label{eqn:rfb}
\RFB\bigl(\ptUps,\rapUps>0\bigr) = \frac{({\rd^2\sigma(\ptUps, \rapUps)}/{\rd \ptUps \rd \rapUps})}{({\rd^2\sigma(\ptUps, -\rapUps)}/{\rd \ptUps \rd \rapUps})},
\end{equation}
where \rapUps is positive. 
We measure event activity near the measured \PgU meson using the number of reconstructed tracks, \Ntracks, in the region $\abs{\etalab} < 2.4$ 
(a detailed discussion of the event-activity variables can be found in Ref.~\cite{HIN-13-003}).
To measure event activity further from the \PgU meson, we use the sum of deposited transverse energy \ET in $4 < \abs{\etalab} < 5.2$.
Figure~\ref{fig:RFBvsHFNtracksInt} shows the \RFB as a function \Ntracks (\cmsLeft), and \ET (\cmsRight).
The uncorrected mean values of the event activity variables in minimum bias \pPb collisions are $\langle \Ntracks \rangle = 41$ and $\langle \ET \rangle = 14.7\GeV$. 
The measured \RFB remains consistent with unity at all levels of event activity for all three \PgU states.
This observation is independent of the $\eta$ region used to measure event activity.
The ALICE Collaboration determined a value of \RFB consistent with unity for \PgUa for integrated event activity for \PgU mesons in the forward ($2.03 < \rapUps < 3.53$) and backward ($-4.46 < \rapUps < -2.96$) rapidity regions~\cite{ALICERpA}. 
The LHCb Collaboration also measured \PgUa \RFB in the forward ($1.5 < \rapUps < 4.0$) and backward ($-5.0 < \rapUps < -2.5$) rapidity regions and reported an integrated \RFB of slightly less than unity~\cite{LHCbRpA}.
In contrast to \PgU results reported here, the \RFB for prompt and nonprompt \JPsi\ were found by CMS to decrease with increasing \ET~\cite{HIN-14-009}.

\begin{figure}[tbh]
\centering
\includegraphics[width=0.48\textwidth]{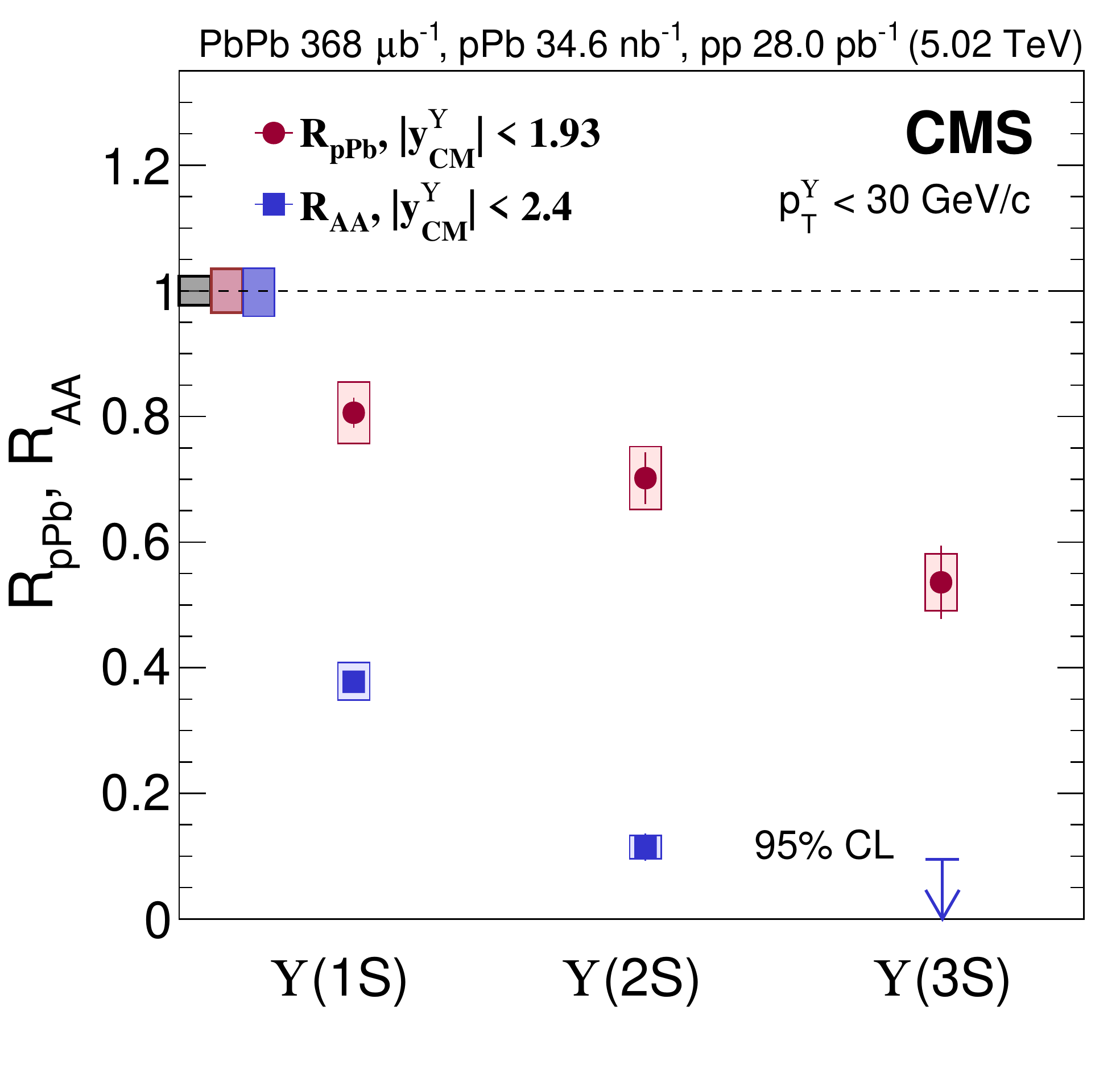}
\caption{\RpPb of \PgUa, \PgUb and \PgUc (red circles) for the integrated kinematic range $0<\ptUps<30\GeVc$ and $\abs{\rapUps}<1.93$. The \RpPb results are compared to the CMS results on \YnS \RAA (blue squares for \PgUa and \PgUb and blue arrow for the upper limit at 95\% confidence level (\CL) on \PgUc) for $0<\ptUps<30\GeVc$ and $\abs{\rapUps}<2.4$, at the same energy~\cite{HIN-16-023}. Vertical bars represent statistical and fit uncertainties and filled boxes around points represent systematic uncertainties. The gray and red boxes around the line at unity depict the uncertainty in the \pp and \pPb luminosity normalizations (2.3 and 3.5\%), respectively. The blue box around unity depicts the global uncertainty pertaining to \PbPb data ($^{+3.6\%}_{-4.1\%}$)~\cite{HIN-16-023}.}
\label{fig:RpAint}
\end{figure}

Figure~\ref{fig:RpAint} shows the integrated \RpPb of \PgU states as well as the \RAA observed in \PbPb collisions~\cite{HIN-16-023} at $\sqrtsNN = 5.02\TeV$. The 95\% confidence level upper limit on the \PgUc \RAA is depicted using an arrow. 
The data indicate an ordering of nuclear modification for the \PgU family with $\RpPb(1S) > \RpPb(2S) > \RpPb(3S):$
\begin{linenomath}
\begin{equation*}
\label{eqn:results}
  \begin{aligned}
  \RpPb(\PgUa) &= 0.806 \pm 0.024 \stat \pm 0.059 \syst, \\
  \RpPb(\PgUb) &= 0.702 \pm 0.041 \stat \pm 0.058 \syst, \\
  \RpPb(\PgUc) &= 0.536 \pm 0.058 \stat \pm 0.050 \syst. \\
  \end{aligned}
\end{equation*}
\end{linenomath}
We determine the \textit{p}-value of the observed suppression of \PgU states in \pPb relative to \pp collisions against the hypothesis of $A$-scaling, which predicts no nuclear modification.
Given the uncertainties in the measured \RpPb values for \PgUa, \PgUb, and \PgUc, we determine these \textit{p}-values to be $1.16\ten{-3}$, $1.36\ten{-5}$, and $6.84\ten{-10}$, respectively.

Given that initial-state CNM models predict equal nuclear modification to all three \PgU states in contrast to final-state CNM, which result in different levels of nuclear modification, 
we can determine the \textit{p}-value of the observed additional suppression of each excited state compared to the ground state. 
This can be done under the hypothesis that no final-state CNM effects are evident, and the \RpPb of the excited and ground states are equal.
The $p$-values of the measured lower \RpPb values of the excited states relative to \PgUa are $1.24\ten{-1}$ for \PgUb and $1.02\ten{-3}$ for \PgUc, corresponding to significances of 1.2 and 3.1 standard deviations, respectively.

Figure~\ref{fig:RpAint} illustrates that the measured modifications in \YnS production in \pPb collisions are considerably smaller than those seen in \PbPb collisions~\cite{HIN-16-023}.
A direct comparison of the \RAA to the \RpPb requires model-dependent scaling of the \RpPb to reflect modification by two lead nuclei in \PbPb collisions instead of one.
Such a comparison of the observed modification effect of CNM on bottomonia in \pPb to the nucleus-nucleus collision environment is needed to determine if hot nuclear matter effects in the QGP result in additional suppression of bottomonia in \PbPb. 
Additional modification in \PbPb compared to \pPb collisions is expected from the presence of color deconfinement as predicted by Refs.~\cite{brambilla,harris,andronic,emerick,gunion}, and by larger comover interaction effects in the dense medium~\cite{FerreiroComp}.

Tabulated results are provided in the HEPData record for this analysis~\cite{hepdata}.

\section{Summary}
The \YnS (where $n=1$, 2, 3) family is studied in proton-lead (\pPb) collisions at $\sqrtsNN=5.02\TeV$ and the production cross sections are presented. Using pp collision data obtained at the same collision energy, 
the nuclear modification factors \RpPb in \pPb collisions for the three \PgU states are measured. 
Compared to the hypothesis of scaling by the number of nucleons $A$, we find the \YnS yields to be suppressed.
This suppression is observed over the entire kinematic range that is studied, \ie, transverse momentum $\ptUps<30\GeVc$ and center-of-mass rapidity $\abs{\rapUps}<1.93$.
The suppression level is constant both as a function of \ptUps\ and of \rapUps within the experimental uncertainties.
An indication of higher separation of the excited states with $\ptUps<6\GeVc$ is observed in the Pb-going direction.

The forward-backward production ratios \RFB of \YnS states are studied 
as a function of event activity in two regions:
A midrapidity region (where the \YnS states were measured), and a region with at least two units of rapidity separation from any measured \YnS state.
The \RFB values are consistent with unity for all states, independent of the region used to measure the event activity. 

The integrated nuclear modification factors for \YnS in \pPb collisions are compared with those measured in \PbPb collisions.
The nuclear modification factors \RAA in \PbPb collisions are much smaller than the corresponding \RpPb value for each state.
However, a similar ordering of the measured \RpPb(\YnS) is observed, with \PgUa the least suppressed.
This suggests the presence of final-state effects in \pPb collisions, consistent with predictions from models that break up the bound quarkonium states via interactions with comoving particles from the underlying event.
These results will help us to understand how bottomonia are modified in heavy-ion collisions.

\begin{acknowledgments}
  We congratulate our colleagues in the CERN accelerator departments for the excellent performance of the LHC and thank the technical and administrative staffs at CERN and at other CMS institutes for their contributions to the success of the CMS effort. In addition, we gratefully acknowledge the computing centers and personnel of the Worldwide LHC Computing Grid and other centers for delivering so effectively the computing infrastructure essential to our analyses. Finally, we acknowledge the enduring support for the construction and operation of the LHC, the CMS detector, and the supporting computing infrastructure provided by the following funding agencies: BMBWF and FWF (Austria); FNRS and FWO (Belgium); CNPq, CAPES, FAPERJ, FAPERGS, and FAPESP (Brazil); MES and BNSF (Bulgaria); CERN; CAS, MoST, and NSFC (China); MINCIENCIAS (Colombia); MSES and CSF (Croatia); RIF (Cyprus); SENESCYT (Ecuador); MoER, ERC PUT and ERDF (Estonia); Academy of Finland, MEC, and HIP (Finland); CEA and CNRS/IN2P3 (France); BMBF, DFG, and HGF (Germany); GSRI (Greece); NKFIA (Hungary); DAE and DST (India); IPM (Iran); SFI (Ireland); INFN (Italy); MSIP and NRF (Republic of Korea); MES (Latvia); LAS (Lithuania); MOE and UM (Malaysia); BUAP, CINVESTAV, CONACYT, LNS, SEP, and UASLP-FAI (Mexico); MOS (Montenegro); MBIE (New Zealand); PAEC (Pakistan); MSHE and NSC (Poland); FCT (Portugal); JINR (Dubna); MON, RosAtom, RAS, RFBR, and NRC KI (Russia); MESTD (Serbia); MCIN/AEI and PCTI (Spain); MOSTR (Sri Lanka); Swiss Funding Agencies (Switzerland); MST (Taipei); ThEPCenter, IPST, STAR, and NSTDA (Thailand); TUBITAK and TAEK (Turkey); NASU (Ukraine); STFC (United Kingdom); DOE and NSF (USA).
  
  \hyphenation{Rachada-pisek} Individuals have received support from the Marie-Curie program and the European Research Council and Horizon 2020 Grant, contract Nos.\ 675440, 724704, 752730, 758316, 765710, 824093, 884104, and COST Action CA16108 (European Union); the Leventis Foundation; the Alfred P.\ Sloan Foundation; the Alexander von Humboldt Foundation; the Belgian Federal Science Policy Office; the Fonds pour la Formation \`a la Recherche dans l'Industrie et dans l'Agriculture (FRIA-Belgium); the Agentschap voor Innovatie door Wetenschap en Technologie (IWT-Belgium); the F.R.S.-FNRS and FWO (Belgium) under the ``Excellence of Science -- EOS" -- be.h project n.\ 30820817; the Beijing Municipal Science \& Technology Commission, No. Z191100007219010; the Ministry of Education, Youth and Sports (MEYS) of the Czech Republic; the Deutsche Forschungsgemeinschaft (DFG), under Germany's Excellence Strategy -- EXC 2121 ``Quantum Universe" -- 390833306, and under project number 400140256 - GRK2497; the Lend\"ulet (``Momentum") Program and the J\'anos Bolyai Research Scholarship of the Hungarian Academy of Sciences, the New National Excellence Program \'UNKP, the NKFIA research grants 123842, 123959, 124845, 124850, 125105, 128713, 128786, and 129058 (Hungary); the Council of Science and Industrial Research, India; the Latvian Council of Science; the Ministry of Science and Higher Education and the National Science Center, contracts Opus 2014/15/B/ST2/03998 and 2015/19/B/ST2/02861 (Poland); the Funda\c{c}\~ao para a Ci\^encia e a Tecnologia, grant CEECIND/01334/2018 (Portugal); the National Priorities Research Program by Qatar National Research Fund; the Ministry of Science and Higher Education, projects no. 0723-2020-0041 and no. FSWW-2020-0008 (Russia); MCIN/AEI/10.13039/501100011033, ERDF ``a way of making Europe", and the Programa Estatal de Fomento de la Investigaci{\'o}n Cient{\'i}fica y T{\'e}cnica de Excelencia Mar\'{\i}a de Maeztu, grant MDM-2017-0765 and Programa Severo Ochoa del Principado de Asturias (Spain); the Stavros Niarchos Foundation (Greece); the Rachadapisek Sompot Fund for Postdoctoral Fellowship, Chulalongkorn University and the Chulalongkorn Academic into Its 2nd Century Project Advancement Project (Thailand); the Kavli Foundation; the Nvidia Corporation; the SuperMicro Corporation; the Welch Foundation, contract C-1845; and the Weston Havens Foundation (USA).
\end{acknowledgments}

\bibliography{auto_generated}\cleardoublepage \appendix\section{The CMS Collaboration \label{app:collab}}\begin{sloppypar}\hyphenpenalty=5000\widowpenalty=500\clubpenalty=5000\cmsinstitute{Yerevan~Physics~Institute, Yerevan, Armenia}
A.~Tumasyan
\cmsinstitute{Institut~f\"{u}r~Hochenergiephysik, Vienna, Austria}
W.~Adam\cmsorcid{0000-0001-9099-4341}, F.~Ambrogi\cmsorcid{0000-0002-9486-0444}, T.~Bergauer\cmsorcid{0000-0002-5786-0293}, M.~Dragicevic\cmsorcid{0000-0003-1967-6783}, J.~Er\"{o}, A.~Escalante~Del~Valle\cmsorcid{0000-0002-9702-6359}, M.~Flechl\cmsorcid{0000-0003-4750-3548}, R.~Fr\"{u}hwirth\cmsAuthorMark{1}, M.~Jeitler\cmsAuthorMark{1}\cmsorcid{0000-0002-5141-9560}, N.~Krammer, I.~Kr\"{a}tschmer\cmsorcid{0000-0002-5636-9259}, D.~Liko, T.~Madlener\cmsorcid{0000-0002-0128-6536}, I.~Mikulec, N.~Rad, J.~Schieck\cmsAuthorMark{1}\cmsorcid{0000-0002-1058-8093}, R.~Sch\"{o}fbeck\cmsorcid{0000-0002-2332-8784}, M.~Spanring\cmsorcid{0000-0001-6328-7887}, W.~Waltenberger\cmsorcid{0000-0002-6215-7228}, C.-E.~Wulz\cmsAuthorMark{1}\cmsorcid{0000-0001-9226-5812}, M.~Zarucki\cmsorcid{0000-0003-1510-5772}
% ...
\vskip\cmsinstskip
% ...
1:~Also at TU Wien, Wien, Austria\\
% ...
\end{sloppypar}
\end{document}